\documentclass[a4paper,11pt]{article}
\usepackage{jheppub,xcolor}
\newcommand{\TeV}{\text{TeV}}
\newcommand{\GeV}{\text{GeV}}
\newcommand{\MeV}{\text{MeV}}
\newcommand{\ifb}{\text{fb}^{-1}}
\newcommand{\iab}{\text{ab}^{-1}}

\newcommand{\eff}{\text{eff}}
\newcommand{\SM}{\text{SM}}
\newcommand{\LP}{\text{LP}}

\title{\boldmath Gravitational waves from first order electroweak phase transition in models with the $U(1)_X^{}$ gauge symmetry}

\preprint{UT-HET-123,OU-HET-957,KIAS-P18013}

\author[a,b]{Katsuya Hashino,}
\author[a]{Mitsuru Kakizaki,}
\author[b]{Shinya Kanemura,}
\author[c]{Pyungwon Ko}
\author[c]{and Toshinori Matsui}

\affiliation[a]{Department of Physics, University of Toyama, \\
3190 Gofuku, Toyama 930-8555, Japan}
\affiliation[b]{Department of Physics, Osaka University, \\
Toyonaka, Osaka 560-0043, Japan}
\affiliation[c]{School of Physics, Korea Institute for Advanced Study, \\
85 Hoegiro, Dongdaemun-gu, Seoul 02455, Republic of Korea}

\emailAdd{hashino@het.phys.sci.osaka-u.ac.jp}
\emailAdd{kakizaki@sci.u-toyama.ac.jp}
\emailAdd{kanemu@het.phys.sci.osaka-u.ac.jp}
\emailAdd{pko@kias.re.kr}
\emailAdd{matsui@kias.re.kr}

\abstract{
 We consider a standard model extension equipped with a dark sector where the $U(1)_X^{}$ Abelian gauge symmetry is spontaneously broken by the dark Higgs mechanism.
 In this framework, we investigate patterns of the electroweak phase transition as well as those of the dark phase transition, and examine detectability of gravitational waves (GWs) generated by such strongly first order phase transition.  
 It is pointed out that the collider bounds on the properties of the discovered Higgs boson exclude a part of parameter space that could otherwise generate detectable GWs.
 After imposing various constraints on this model, it is shown that GWs produced by multi-step phase transitions are detectable at future space-based interferometers, such as LISA and DECIGO, if the dark photon is heavier than 25 GeV. 
 Furthermore, we discuss the complementarity of dark photon searches or dark matter searches with the GW observations in these models with the dark gauge symmetry.
}

\begin{document} 
\maketitle
\flushbottom

\section{Introduction}
\label{sec:intro}

 Due to the discovery of the Higgs boson with a mass of $m_h^{} \simeq 125~\GeV$~\cite{Aad:2012tfa,Chatrchyan:2012xdj}, the standard model (SM) has been experimentally established as an effective theory that describes spontaneous breaking of the electroweak (EW) symmetry.
 However, the properties of the discovered Higgs boson as well as the dynamics behind the electroweak symmetry breaking (EWSB) are still unknown.
 In addition, there exist phenomena that require new physics beyond the SM (BSM), such as neutrino oscillations, baryon asymmetry of the Universe, the existence of dark matter (DM) and inflation.
 One of the most intriguing ideas is to relate new physics that accounts for these phenomena with physics in the Higgs sector.

 One of the interesting phenomena that relate such new physics and the Higgs sector is electroweak baryogensis (EWBG)~\cite{Kuzmin:1985mm}.
 For generating baryon asymmetry, one must satisfy Sakharov's conditions: Violation of baryon number; simultaneous violation of $C$ and $CP$ symmetry; and departure from thermal equilibrium~\cite{Sakharov:1967dj}.
 EWBG scenarios can be realized by extending the Higgs sector since the first condition is provided by the sphaleron process, the second by additional CP violating phases other than the Kobayashi-Maskawa phase and the third by strongly first order phase transition (1stOPT) (see e.g. Refs.~\cite{Quiros:1994dr,Trodden:1998ym,Riotto:1998bt}).

 There are several ways to realize strongly 1stOPT.
 One approach is to invoke the non-decoupling thermal loop effects, which modify the finite temperature effective potential~\cite{Funakubo:1993jg,Cline:1995dg,Cline:1996mga,Kanemura:2004ch,Espinosa:2007qk,Noble:2007kk,AKS,Kanemura:2011fy,Gil:2012ya,Tamarit:2014dua,Kanemura:2014cka,Blinov:2015vma,Fuyuto:2015vna,Hashino:2015nxa,Kakizaki:2015wua,Hashino:2016rvx,Dorsch:2016nrg,Basler:2016obg,Marzola:2017jzl}. 
 In this case, electroweak phase transition (EWPT) occurs through one-step phase transition (PT) along the SM Higgs field direction.
 At the same time, these additional scalar loop effects affect the effective potential at zero temperature, leading to significant deviation in the triple Higgs boson coupling ($hhh$ coupling) from the SM value typically by larger than $10\%$~\cite{Kanemura:2004ch,Noble:2007kk,AKS,Kanemura:2011fy,Tamarit:2014dua,Hashino:2015nxa,Kanemura:2014cka,Kakizaki:2015wua,Hashino:2016rvx}. 
 It should be noticed that the one-step PT is not the only possibility if one considers extended Higgs sectors. 
 In models where the SM Higgs boson mixes with additional Higgs bosons, multi-step PTs which involves the first order EWPT along the SM Higgs field direction can be realized~\cite{Pietroni:1992in,Apreda:2001us,Menon:2004wv,Funakubo:2005pu,Profumo:2007wc,Ashoorioon:2009nf,Espinosa:2011ax,Chung:2011it,Carena:2011jy,Huang:2014ifa,Fuyuto:2014yia,Kotwal:2016tex,Profumo:2014opa,Tenkanen:2016idg,Huang:2016cjm,Hashino:2016xoj,Bian:2017wfv,Chen:2017qcz,Chiang:2017nmu}.
 Multi-step PTs are possible also in models with additional symmetries such as global $Z_2$ symmetry~\cite{Cline:2012hg,Curtin:2014jma,Vaskonen:2016yiu,Curtin:2016urg,Chao:2017vrq,Beniwal:2017eik,Kurup:2017dzf,Jain:2017sqm}, global $Z_3$ symmetry~\cite{Kang:2017mkl} and gauge symmetry~\cite{Chao:2014ina}.
 In a class of models where the strongly 1stOPT is caused by the field mixing, the predicted values for the Higgs boson couplings with the weak gauge bosons and with the SM fermions can deviate significantly from their SM predictions~\cite{Fuyuto:2014yia,Hashino:2016xoj}.
 There are also nightmare scenarios where multi-step PT occurs even for weak couplings~\cite{Ashoorioon:2009nf,Curtin:2014jma} and leaves no measurable footprint at collider experiments.
  
 We emphasize that the nature of EWPT can be probed by exploring the Higgs sector at ongoing and future experiments.
 Models predicting significant deviations in various Higgs boson couplings can be tested at the LHC~\cite{CMS:2013xfa} as well as at future lepton colliders including, the International Linear Collider (ILC)~\cite{ILC}, the Compact Linear Collider (CLIC)~\cite{CLIC} and the Future Circular Collider of electrons and positrons (FCC-ee)~\cite{FCC-ee}.
 As for the $hhh$ coupling, the high-luminosity LHC will constrain the deviation up to a factor~\cite{LHChhh1,LHChhh2}.
 If the International Linear Collider (ILC) with $\sqrt{s}=1~{\rm TeV}$ is realized the $hhh$ coupling will be determined with $10\%$ accuracy~\cite{ILCHiggsWhitePaper,Moortgat-Picka:2015yla,Fujii:2015jha}~\footnote{
 Recently, a new plan for the ILC project was proposed~\cite{Fujii:2017vwa,Asai:2017pwp}, in which the collision energy is set to $250~\GeV$ and the operation is for about 10 years, accumulating the integrated luminosity up to $2~\iab$.
 Nevertheless, one can expect that the nominal $1~\TeV$ collision energy might be realized, depending on our findings in future.
}.
 The capability of measuring the $hhh$ coupling at future hadron colliders with $\sqrt{s} = 100~{\rm TeV}$ has been discussed~\cite{He:2015spf}. 

 On the cosmological side, the strongly 1stOPT that occurs in the early Universe produces stochastic GWs detectable at future space-based interferometers~\cite{Kosowsky:1991ua,Kamionkowski:1993fg,Kosowsky:2001xp,Dolgov:2002ra,Apreda:2001us,Grojean:2006bp,Espinosa:2008kw,Kehayias:2009tn,Kakizaki:2015wua,Caprini:2015zlo,Huber:2015znp,Hashino:2016rvx,Dev:2016feu,Chala:2016ykx,Kobakhidze:2016mch,Addazi:2016fbj}.  
 Until now, several GWs events generated by the mergers of binary black holes and binary neutron stars have been observed at the Advanced LIGO and Advanced VIRGO~\cite{aLIGO}. 
 The worldwide network of GW detectors including Advanced LIGO~\cite{Harry:2010zz}, Advanced VIRGO~\cite{Accadia:2009zz} and KAGRA~\cite{Somiya:2011np} will reveal astronomical problems.
 In future, planned space-based interferometers such as LISA~\cite{Seoane:2013qna}, DECIGO~\cite{Kawamura:2011zz} and BBO~\cite{Corbin:2005ny} will survey GWs in the millihertz to decihertz range, which is the typical frequency of GWs from the first order EWPT.
 Even the above-mentioned nightmare scenarios can be investigated by measuring GWs at these future interferometers~\cite{Ashoorioon:2009nf,Kang:2017mkl}.
 Therefore, using the synergy between the measurement of the Higgs boson couplings at colliders and the observation of GWs at interferometers, one can scrutinize the nature of EWPT and distinguish EWBG scenarios.

 Among various extensions of the Higgs sector, models with Higgs portal DM can account for the relic abundance of DM, and thus have been extensively studied in recent years. 
 
 For example, Higgs portal DM within the effective field theory (EFT) framework has been investigated in Refs.~\cite{Kanemura:2010sh,Djouadi:2011aa,Beniwal:2015sdl,Ko:2016xwd}.  
 The simplest gauge invariant renormalizable Higgs portal DM model for the singlet scalar DM (SSDM) has been investigated in Refs.~\cite{Gonderinger:2009jp,Cline:2012hg,Curtin:2014jma,Vaskonen:2016yiu,Chao:2017vrq,Beniwal:2017eik,Kurup:2017dzf,Jain:2017sqm}.
 As for the singlet fermion and vector DM cases with Higgs portal interactions, the EFT description has some drawbacks due to the nonrenormalizability of the Higgs portal interaction and violation of gauge symmetry, respectively. 
 These problems can be resolved by introducing a singlet scalar field for the singlet fermion DM (SFDM) models~\cite{Baek:2011aa,Baek:2012uj,Fairbairn:2013uta,Li:2014wia,Ettefaghi:2017vbh} and a dark Higgs field in the vector DM (VDM) models~\cite{Lebedev:2011iq,Farzan:2012hh,Baek:2012se,Chao:2014ina,Duch:2015jta}.
 PT (and GW production) in Higgs portal DM models has been studied for the cases of SSDM~\cite{Cline:2012hg,Curtin:2014jma,Vaskonen:2016yiu,Chao:2017vrq,Beniwal:2017eik,Kurup:2017dzf,Jain:2017sqm}(\cite{Huang:2016cjm}), SFDM~\cite{Li:2014wia} and VDM~\cite{Chao:2014ina}.
 
 In this paper, we shall focus on a model with gauged dark $U(1)_X^{}$ symmetry as one of the viable Higgs portal DM models.
 First, we address a more general case where the dark photon decays into SM particles through the $U(1)$ gauge kinetic mixing term and investigate the nature of EWPT and dark PT.
 We revisit the complementarity of dark photon searches and GW observations in this model with strongly 1stOPT~\cite{Addazi:2017gpt}, and find the lower bound of $m_X$ in the light of current collider bounds.
 We then explore the Higgs portal DM model with VDM, which is stabilized by introducing a discrete $Z_2$, and investigate the complementarity of the detection of GWs from the strongly 1stOPT, collider bounds and DM searches.
 
 This paper is organized as follows.
 In Sec.~\ref{sec:model}, we briefly review the model with the $U(1)_X^{}$ dark gauge symmetry and show formulae about the properties of the Higgs bosons and the finite temperature
 effective potential.
 We discuss PT patterns at finite temperature and introduce quantities which characterize the spectrum of GWs produced by bubble collisions in Sec.~\ref{sec:ewpt}.
 In Sec.~\ref{sec:results}, our numerical results about the prospect of the detection of GWs are shown for various benchmark points after imposing theoretical and experimental constraints.
 Sec.~\ref{sec:conclusions} is devoted to discussion and conclusions.
 The tree-level unitarity of the Higgs self-couplings is discussed using analytic formulae in Appendix~\ref{sec:analytic}.
 The one-loop renormalization group equations for the model parameters are given in Appendix~\ref{sec:beta}.
 Constraints on the DM properties are discussed in Appendix~\ref{sec:dm}.

\section{Model with dark $U(1)_X$  gauge symmetry}
\label{sec:model}

 We consider a model with a dark sector where the $U(1)_X^{}$ Abelian gauge symmetry is spontaneously broken by the so-called dark Higgs mechanism. 
 We introduce a complex scalar (called dark Higgs boson) $S$ with $U(1)_X^{}$-charge $Q_X^{}$ and the $U(1)_X^{}$ gauge field (dark photon) $X_\mu^0$. 
 In generic, there appears the gauge kinetic mixing term between the $U(1)_X^{}$ gauge boson $X_\mu^0$ and the hypercharge $U(1)_Y^{}$ gauge boson $B_\mu^{}$~\cite{Holdom:1985ag}, and the Lagrangian for the newly introduced fields is (e.g. Ref.~\cite{Addazi:2017gpt})
\begin{align}
{\cal L} = - \frac{1}{4} X_{\mu\nu} X^{\mu\nu} - \frac{\epsilon}{2} X_{\mu \nu} B^{\mu \nu} + |D_\mu S|^2 - V_0(\Phi, S)
\label{eq:lagrangian}
\end{align}
where $X_{\mu\nu}=\partial_\mu X_\nu^0 - \partial_\nu X_\mu^0$ and
$B_{\mu\nu}^0=\partial_\mu B_{\nu}^0 - \partial_\nu B_{\mu}^0$, \
 and the covariant derivative is defined as $D_\mu = \partial_\mu + i g_X Q_X X_\mu^0$.
Here, the Higgs potential is given by 
\begin{align}
V_0(\Phi, S) =
-\mu_\Phi^2|\Phi|^2
-\mu_S^2 |S|^2
+\lambda_\Phi^{} |\Phi|^4
+\lambda_S^{} |S|^4
+\lambda_{\Phi S}^{} |\Phi|^2 |S|^2.
\label{eq:full_theory}
\end{align}
 We normalize the $U(1)_X^{}$ charge of $S$, $Q_S^{} \equiv Q_X^{}(S)$, as $Q_S^{}=1$.
 Since the viable parameter range for $\epsilon$ is too small to affect PT~\cite{Addazi:2017gpt}, we focus on the rest six parameters, i.e. $\mu^2_\Phi, \mu^2_S, \lambda_\Phi^{}, \lambda_S^{}$, $\lambda_{\Phi S}^{}$ and $g_X^{}$.
 We refer to the model with the kinetic mixing as ``Model A'' in this paper.
 In this paper, we focus on the cases where $\mu_\Phi$ and $\mu_S$ are of the same order since we address the complementarity between collider experiments and cosmological observations in exploring new physics at around the EW scale. In the large $\mu_S$ limit, the singlet field decouples from the SM. GW production from such high-scale phase transition is discussed in e.g. Refs.~\cite{Dev:2016feu,Jinno:2015doa,Jaeckel:2016jlh,Jinno:2016knw,Balazs:2016tbi}.

 If we introduce a  discrete $Z_2$ symmetry under which the vector boson $X_\mu^0$ is odd, the kinetic mixing term $X_{\mu \nu} B^{\mu \nu}$ is prohibited, stabilizing  $X_\mu^0$.
 In this case, the vector boson $X_\mu^0$ can be an excellent candidate for DM~\cite{Lebedev:2011iq,Farzan:2012hh,Baek:2012se,Chao:2014ina,Duch:2015jta}~\footnote{
 Without imposing $Z_2$ symmetry, $X_\mu^0$ can be a DM candidate with long life time which requires $m_X \lesssim {\cal O}(\MeV)$ to avoid that $X_\mu^0$ decays to leptons~\cite{Jaeckel:2013ija}.
 However, we do not need to care such case in the context of this paper, because we have not found any points of 1stOPT as shown later.
}. 
 The case without the kinetic mixing is referred to as "Model B".
 As far as PT is concerned, Model A and Model B can be discussed on the same footing.

 After the EWSB, the two Higgs multiplets can be expanded as
\begin{align}
\Phi=
\left(
 \begin{array}{c}
  w^+ \\ \frac{1}{\sqrt{2}}(v_\Phi+\phi_\Phi+i z^0)
 \end{array}
\right), \quad
S=\frac{1}{\sqrt{2}}(v_S+\phi_S+i x^0), 
\end{align}
where $v_\Phi$ $(= 246~\GeV)$ and $v_S$ are the corresponding vacuum expectation values (VEV), $\phi_\Phi$ and $\phi_S$ are physical degrees of freedom which mix with each other through the $\lambda_{\Phi S}$ term in the Higgs potential (Eq.~(2)). 
 The Nambu-Goldstone modes $w^\pm$, $z^0$ and $x^0$ are absorbed by the gauge bosons $W^\pm_\mu$, $Z^0_\mu$ and $X_\mu^0$.
 The mass of $X_\mu^0$ is $m_{X} = g_X |Q_S| v_S$ (see also Ref.~\cite{Farzan:2012hh}).

 The phase structure of our model is analyzed in the classical field space spanned by $\langle \Phi \rangle = (0, \varphi_\Phi^{}/\sqrt{2})$ and $\langle S \rangle = \varphi_S^{}/\sqrt{2}$. 
 The Higgs potential is modified from its tree-level form due to radiative corrections.
 At zero temperature, the effective potential at the one-loop level is given by~\cite{Coleman:1973jx}
\begin{align}
  V_{\eff, T=0}^{}(\varphi_\Phi^{},\varphi_S^{})
  =V_0(\varphi_\Phi^{},\varphi_S^{}) +\sum_i n_i^{} \ 
  \frac{M^4_i(\varphi_\Phi^{},\varphi_S^{})}{64\pi^2}
  \left(\ln\frac{M^2_i(\varphi_\Phi^{},\varphi_S^{})}{Q^2} -c_i \right),
\end{align}
where $Q$ is the renormalization scale, which is set at $v_\Phi^{}$ in our analysis~\footnote{
 Although we usually determine the $Q$ by the renormalization conditions at $T=0$, it is technically complicated for models with nonzero VEVs in addition to the $v_\Phi^{}$.
 As a reference, effects of renormalization group running at the critical temperature $T_c^{}$ are discussed in Ref.~\cite{Chiang:2017nmu}.
}.
 Here, $n_i$ and $M_i(\varphi_\Phi,\varphi_S)$ stand for the degrees of the freedom and the field-dependent masses for particles $i=h, H, w^\pm, z^0, x^0, W_{\mu\,(T, L)}^{\pm}, Z_{\mu\,(T, L)}^{0}, X_{\mu\,(T, L)}^{0}, \gamma_{\mu\,(T, L)}^{0}, t$ and $b$, respectively. 
 We adopt the mass-independent $\overline{\rm MS}$ scheme, where the numerical constants $c_i$ are set at $3/2$ ($5/6$) for scalars and fermions (gauge bosons).
 We impose the conditions that the tadpole terms at the one-loop level vanish as
\begin{align}
\left\langle \frac{\partial V_{\eff, T=0}^{}}{\partial \varphi_\alpha^{}} \right\rangle=0, 
\label{eq:tad}
\end{align}
for $\alpha=\Phi$ and $S$.
 The angle bracket $\langle\cdots\rangle$ denotes the corresponding field-dependent value evaluated at our true vacuum $(\varphi_\Phi^{}, \varphi_S^{}) = (v_\Phi^{}, v_S^{})$.

 The interaction basis states $\phi_\Phi^{}$ and $\phi_S^{}$ are relations with their mass eigenstates $h$ and $H$ through 
\begin{align}
\left(
 \begin{array}{c}
  \phi_\Phi \\ \phi_S
 \end{array}
\right)
=
\left(
 \begin{array}{cc}
  c_\theta & -s_\theta \\
  s_\theta & c_\theta
 \end{array}
\right)
\left(
 \begin{array}{c}
  h \\ H
 \end{array}
\right) ,
\end{align}
with $c_\theta\equiv\cos\theta$, $s_\theta\equiv\sin\theta$.
 The one-loop improved mass squared matrix of the real scalar bosons in the $(\phi_\Phi, \phi_S)$ basis is then diagonalized as
\begin{align}
  m_{\alpha \beta}^2 = \left\langle 
  \frac{\partial^2 V_{\eff, T=0}^{}}{\partial \varphi_\alpha^{} \partial \varphi_\beta^{}}
  \right\rangle
=
\begin{pmatrix}
c_\theta & -s_\theta \\
s_\theta & c_\theta
\end{pmatrix}
\begin{pmatrix}
m_{h}^2 & 0 \\
0 & m_{H}^2
\end{pmatrix}
\begin{pmatrix}
c_\theta & s_\theta \\
-s_\theta & c_\theta
\end{pmatrix}.
\label{eq:mass}
\end{align}
 We denote $h$ and $H$ as the discovered Higgs boson with the mass $m_h^{}=125~\GeV$ and the additional neutral Higgs boson with mass eigenvalue $m_{H}$, so that the absolute value of the mixing angle $|\theta^\circ|$ is less than $45^\circ$.
 In our analysis, we regard $v_\Phi^{}$, $m_h^{}$, $m_H^{}$, $\theta$, $g_X$ and $m_X$ as input parameters, and $\mu_\Phi^2$, $\mu_S^2$, $\lambda_\Phi^{}$, $\lambda_S^{}$, $\lambda_{\Phi S}^{}$ and $v_S^{}$ as derived parameters from them.

 The tree-level interactions of $h$ and $H$ with the SM gauge bosons $V$(=$W_\mu^\pm$, $Z_\mu^0$) and with the SM fermions  $F$ are given by
\begin{align}
{\cal L}_{\Phi V V} &= \frac{1}{v_\Phi} (h c_\theta-H s_\theta)(2 m_W^2 W^+_\mu W^{- \mu}+m_Z ^2 Z^0_\mu Z^{0 \mu}) \\
{\cal L}_{\rm Yukawa} &= -\sum_F\frac{m_F}{v_\Phi} (h c_\theta-H s_\theta)   \bar{F} F,
\end{align}
respectively. 
 These Higgs boson couplings in our model normalized by the corresponding SM ones are universally given by
\begin{align}
\kappa \equiv 
\frac{g_{h VV}}{g_{h VV}^\SM} =\frac{g_{h FF}}{g_{h FF}^\SM}=c_\theta. 
\end{align}

 Using the effective potential approach, the $hhh$ coupling is computed as 
 \begin{align}
\lambda_{hhh}^{}&=
c_\theta^3\left\langle\frac{\partial^3 V_{\eff, T=0}}{\partial \varphi_\Phi^3}\right\rangle
+c_\theta^2s_\theta\left\langle\frac{\partial^3 V_{\eff, T=0}}{\partial \varphi_\Phi^2\partial \varphi_S}\right\rangle
+c_\theta s_\theta^2\left\langle\frac{\partial^3 V_{\eff, T=0}}{\partial \varphi_\Phi\partial \varphi_S^2}\right\rangle
+s_\theta^3\left\langle\frac{\partial^3 V_{\eff, T=0}}{\partial \varphi_S^3}\right\rangle. 
\end{align}
 Its SM prediction is approximately given by~\cite{Kanemura:2002vm,Kanemura:2004mg}
\begin{align}
  \lambda_{hhh}^\SM  \simeq \frac{3m_{h}^2}{v_\Phi}
  \left[
	1-\frac{1}{\pi^2}\frac{m_t^4}{v_\Phi^2m_{h}^2}
\right].
\end{align}
 The deviation in the $hhh$ coupling is defined as
\begin{align}
\frac{\Delta \lambda_{hhh}^{}}{\lambda_{hhh}^\SM}
\equiv \frac{\lambda_{hhh}^{}-\lambda_{hhh}^\SM }{\lambda_{hhh}^\SM}.
\label{Eq.hhh}
\end{align}

 At finite temperatures, the effective potential receives additional contributions from thermal loop diagrams, and is modified to~\cite{Dolan:1973qd}
\begin{align}
  V_{\eff,T}^{}[M_i^2(\varphi_\Phi^{},\varphi_S^{})]
  =V_{\eff, T=0}^{}(\varphi_\Phi^{},\varphi_S^{})
  + \sum_i n_i \ \frac{T^4}{2\pi^2}I_{B,F} 
  \left( \frac{ M^2_i(\varphi_\Phi^{},\varphi_S^{})}{T^2} \right),
\label{eq:effpot_finite}
\end{align}
where $I_{B,F}(a^2)= \int^{\infty}_0 dx \ x^2 \ln \left[1 \mp \exp (-\sqrt{x^2+a^2}) \right]$ for bosons $(-)$  and fermions $(+)$, respectively.
 In order to take ring-diagram contributions into account, we replace the field-dependent masses in the effective potential by~\cite{Carrington:1991hz}
\begin{align}
  M_i^2(\varphi_\Phi^{},\varphi_S^{})
  \to M_i^2(\varphi_\Phi^{},\varphi_S^{}, T)
  = M_i^2(\varphi_\Phi^{},\varphi_S^{})+\Pi_i(T),
\end{align}
where $\Pi_i^{}(T)$ denote the finite temperature contributions to the self energies of the fields $i$.
 The thermally corrected field-dependent masses of the Higgs bosons are
\begin{align}
  M^2_{h, H}(\varphi_\Phi^{},\varphi_S^{},T)
  &=\frac{1}{2}\left(M_{\Phi \Phi}^2+M_{S S}^2 \mp 
    \sqrt{(M_{\Phi \Phi}^2-M_{S S}^2)^2 + 4 M_{\Phi S}^2 M_{S \Phi}^2} \right), \\
  M^2_{w^\pm, z^0}(\varphi_\Phi^{},\varphi_S^{},T)
  &=M_{\Phi \Phi}^2-2\lambda_\Phi^{} \varphi_\Phi^2, \\
  M^2_{x^0}(\varphi_\Phi^{},\varphi_S^{},T)
  &=M_{S S}^2-2\lambda_S^{} \varphi_S^2, 
\end{align}
where
\begin{align}
  \begin{pmatrix}
    M_{\Phi \Phi}^2 & M_{\Phi S}^2 \\
    M_{S \Phi}^2 & M_{S S}^2
  \end{pmatrix}
               = &
  \begin{pmatrix}
    -\mu_\Phi^2 + 3\lambda_\Phi^{} \varphi_\Phi^2
    + \frac{\lambda_{\Phi S}^{}}{2} \varphi_S^2
    & \lambda_{\Phi S}^{} \varphi_\Phi^{} \varphi_S^{} \\
    \lambda_{\Phi S}^{} \varphi_\Phi^{} \varphi_S^{} 
    & -\mu_S^2 + 3\lambda_S^{} \varphi_S^2
    +\frac{\lambda_{\Phi S}^{}}{2}\varphi_\Phi^2
  \end{pmatrix}\nonumber\\
&+\frac{T^2}{48}
	\begin{pmatrix}
	9 g^2+3 g'^2+12 (y_t^2+y_b^2) + 24 \lambda_\Phi^{} 
        + 4 \lambda_{\Phi S}^{} & 0 \\
	0 & 12 g_X^2+16 \lambda_S^{} + 8 \lambda_{\Phi S}^{}
	\end{pmatrix}. \nonumber
\end{align}
 Here, $g$, $g'$ and $g_X$ ($y_t^{}$ and $y_b^{}$) are the gauge couplings of $SU(2)_L$, $U(1)_Y^{}$ and $U(1)_X^{}$ (the top and bottom Yukawa couplings).
 In the $(W^+_\mu, W^-_\mu, W^3_\mu, B^0_\mu)$ basis, the field-dependent masses of the EW gauge bosons are thermally corrected as
\begin{align}
  M^{2(L, T)}_{g}(\varphi, T)
  =
  \frac{\varphi_\Phi^2}{4}\begin{pmatrix} g^2&&& 
    \\ &g^2&& \\ &&g^2&gg' \\ &&gg'&g'^2
  \end{pmatrix} 
  + a^{L, T}_g T^2 \begin{pmatrix} 
  g^2&&& \\ &g^2&& \\ &&g^2& \\ &&&g'^2 \end{pmatrix} 
  \label{eq:gaugethermal}
\end{align}
with $a_{g}^L=11/6$, $a_{g}^T=0$
and that of the $U(1)_X^{}$ gauge boson as
\begin{align}
  M^{2(L, T)}_{X^0}(\varphi_S, T)=
   g_X^2 \varphi_S^2 + a^{L, T}_X g_X^2 T^2, 
  \label{eq:gaugethermal_x}
\end{align}
with $a_{X}^L=1/3$, $a_{X}^T=0$~\cite{Carrington:1991hz,Funakubo:2012qc,Chiang:2017zbz}~\footnote{
 In Ref.~\cite{Chao:2014ina}, the coefficient of the contribution to the longitudinal part of the $X$-boson is $a_{X}^L=2/3$, which does not agree with ours.
}. 
 On the other hand, fermion counterparts do not receive such thermal corrections.
 
 The one-loop effective potential at finite temperature has the notorious problem of gauge dependence,  which has been known for a long time, but no complete treatment has been invented.
 A gauge invariant treatment for evaluating the critical temperature $T_c^{}$ has been discussed in Ref.~\cite{Chiang:2017nmu}. However, the computation of the transition temperature $T_t^{}$, which is relevant to the GW production, requires the high temperature approximation.
 The uncertainties in the prediction of the GW spectrum under specific gauge choices are discussed in Ref.~\cite{Chiang:2017zbz}.
 In this paper, we take the Landau gauge, where the gauge-fixing parameter vanishes $\xi=0$, as a reference although we are aware of the problem pf gauge dependence.

\section{The first order electroweak phase transition}
\label{sec:ewpt}

\begin{figure}[t]
\centering
\includegraphics[width=.42\textwidth]{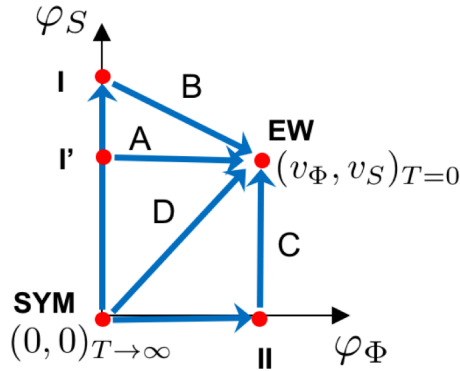}
\vspace{-8mm}
\caption{
Four types of PT path are labeled by A, B, C and D.
Here, SYM and EW denotes the symmetric phase $(\varphi_\Phi^{},\varphi_S
^{})=(0,0)_{T\to \infty}$, and the EW phase $(v_\Phi^{},v_S^{})_{T=0}$.
Intermediate phases are denoted by I $(0,\varphi_S^{}>v_S^{})$, I' $(0,v
_S^{})$ and II $(v_\Phi,0)$.
For details, see Refs.~\cite{Funakubo:2005pu,Profumo:2014opa,Chiang:2017nmu,Vieu:2018nfq}.
 }
\label{fig:PT}
\end{figure}

 As discussed in Refs.~\cite{Funakubo:2005pu,Profumo:2014opa,Chiang:2017nmu}, there are typically four different types of PT path as shown in Fig.~\ref{fig:PT}.
 In our numerical analysis, we impose the condition that the EW phase with massive dark photon ($\varphi_\Phi$, $\varphi_S$)=($v_\Phi$, $v_S$) becomes the global minimum at $T=0$~\cite{Chen:2014ask,Espinosa:2011ax}: 
\begin{align}
V_{\eff, T=0}^{}({\rm EW~phase})<V_{\eff, T=0}^{}({\rm other~phases}).
 \label{eq:PT}
\end{align}

 In order to discuss GWs originating from the first order EWPT in an analytic manner, we introduce several important quantities that parametrize the dynamics of vacuum bubbles following Ref.~\cite{Grojean:2006bp}. 
 The transition temperature $T_t^{}$ is defined such that the bubble nucleation probability per Hubble volume per Hubble time reaches the unity:
\begin{align}
 \left. \frac{\Gamma}{H^4} \right|_{T=T_t^{}}^{}=1.
\end{align}
 The produced GWs are enhanced as the released energy density $\epsilon$ is increased.
 A dimensionless parameter $\alpha$ is defined as the ratio of $\epsilon$ to the radiation energy density $\rho_{\rm rad}^{} =(\pi^2/30) g_*^{} T^4$ at the transition temperature $T_t^{}$:
\begin{align}
  \alpha \equiv \frac{\epsilon(T_t)}{\rho_{\rm rad}(T_t)}.
\end{align}
 For simplicity, the relativistic degrees of freedom is set at $g_*^{}=110.75$, and the temperature dependence of $g_*^{}$ is neglected. 
 The bubble nucleation rate can be parametrized as $\Gamma(t)=\Gamma_0^{}\exp(\beta t)$ at around the transition temperature $T_t^{}$.
 We introduce another dimensionless parameter $\widetilde{\beta}$ as the ration of the inverse of the time variation scale of the bubble nucleation rate $\beta$ to the Hubble parameter at $T=T_t^{}$:
\begin{align}
\widetilde{\beta} \equiv \frac{\beta}{H_t^{}}
 = T_t^{} \frac{d}{d T}\left(\frac{S_3^{}(T)}{T}\right)\Bigg|_{T=T_t}^{},
\end{align}
where $S_3^{}(T)$ is the three-dimensional Euclidean action of the bounce solution of the classical fields that is stretched between the true and false vacua at finite temperature $T$.
 The predicted GW spectrum is expressed in terms of $T_t^{}$, $\alpha$ and $\widetilde{\beta}$. 
 According to Ref.~\cite{sw}, the contribution from sound waves is the main source for stochastic GWs from 1stOPT while those from the bubble wall collision and the turbulence are not significant~\cite{sw}.
 We employ the approximate analytic formula provided in Ref.~\cite{Caprini:2015zlo} for computing the spectrum of the GWs.

\section{Numerical results}
\label{sec:results}

\begin{figure}[t]
\centering
  \includegraphics[width=.49\textwidth]{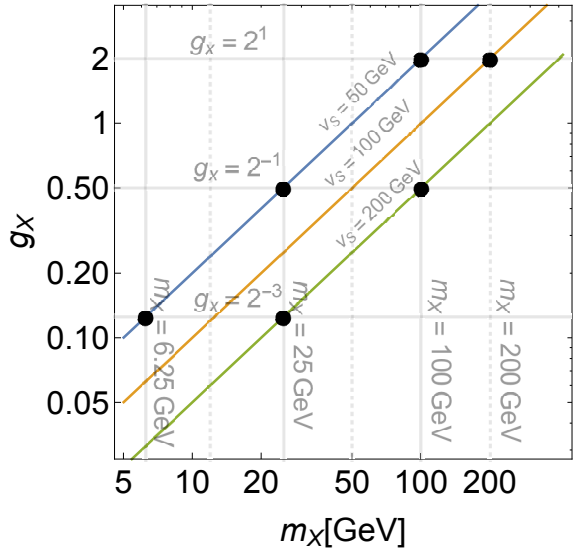}
 \caption{
 The six benchmark points are blobbed on the ($m_X, g_X$)-plane.
 The solid lines show the cases of $v_S [\GeV]=50$ (blue), $100$ (orange), $200$ (green) from the top.
 }
 \label{fig:benchmark}
\end{figure}

 For our numerical analysis of PT, we implement the model introduced in Sec.~\ref{sec:model} into the public code {\tt CosmoTransitions 2.0a2}~\cite{Wainwright:2011kj}, which computes quantities related to cosmological PT in the multi-field space.
 Our analysis is focused on the six benchmark points blobbed in Fig.~\ref{fig:benchmark}. 
 In the following, we detail theoretical and experimental constraints taken into account in our numerical analyses.

 The conditions for vacuum stability for the Higgs potential are given by
\begin{align}
\lambda_\Phi^{}>0, \quad
\lambda_{S}^{}>0, \quad
4\lambda_{\Phi}^{}\lambda_S^{}>\lambda^2_{\Phi S}.
\label{eq:vs}
\end{align}
 By the requirement of perturbative unitarity~\cite{Lee:1977eg}, the magnitudes of the eigenvalues of $S$-wave scattering amplitudes for the longitudinal weak gauge bosons and the scalars must be  smaller than $1/2$, leading to~\cite{Kanemura:2015fra,He:2016sqr}, 
\begin{align}
|\lambda_\Phi|<4 \pi, \quad |\lambda_S|<4 \pi, \quad |\lambda_{\Phi S}|<8 \pi, \quad
3\lambda_\Phi+2\lambda_S + \sqrt{(3\lambda_\Phi-2\lambda_S)^2+2\lambda_{\Phi S}^2}<8 \pi.
\label{eq:perturbativity}
\end{align}
 Further discussions on perturbative unitarity are given in Appendix~\ref{sec:analytic}.

 Electroweak precision measurements constrain parameters in the Higgs sector of our model.
 Since the mass of the discovered Higgs boson is $m_h^{} \simeq 125~\GeV$, the mixing angle of the Higgs bosons is bounded as $\theta \lesssim 23^\circ$ when the mass of the additional Higgs boson is $m_H^{} \gtrsim 400~\GeV$~\cite{Baek:2011aa,Baek:2012uj}.
 The measurements of the Higgs boson decay into weak gauge bosons give constraints on the $hVV$ couplings as $\kappa_Z^{}=1.03^{+0.11}_{-0.11}$ and $\kappa_W^{}=0.91^{+0.10}_{-0.10}$ from the ATLAS and CMS combination of the LHC Run-I data (68\% CL)~\cite{TheATLASandCMSCollaborations:2015bln}.
 In our numerical analysis, we take the 68\% CL bound $\kappa_Z^{}>0.92$ as the lower bound on the mixing angle, namely 
\begin{align}
|\theta| \leq 23.1^\circ.
\label{eq:Hdirect}
\end{align}
 The exclusion limits from the direct searches for the $H$ boson at the LEP and LHC Run-II are examined in Ref.~\cite{Robens:2015gla}.  
 We will show that a large portion of the model parameter space where strongly 1stOPT and detectable GW signals are possible is excluded by the collider bounds on the Higgs bosons discussed above~\footnote{
 The collider constraints on the properties of the Higgs bosons were not considered in Ref.~\cite{Addazi:2017gpt}, where similar models were discussed in the context of GW production from 1stOPT.
}. 

\begin{figure}[t]
\centering
\includegraphics[width=.49\textwidth]{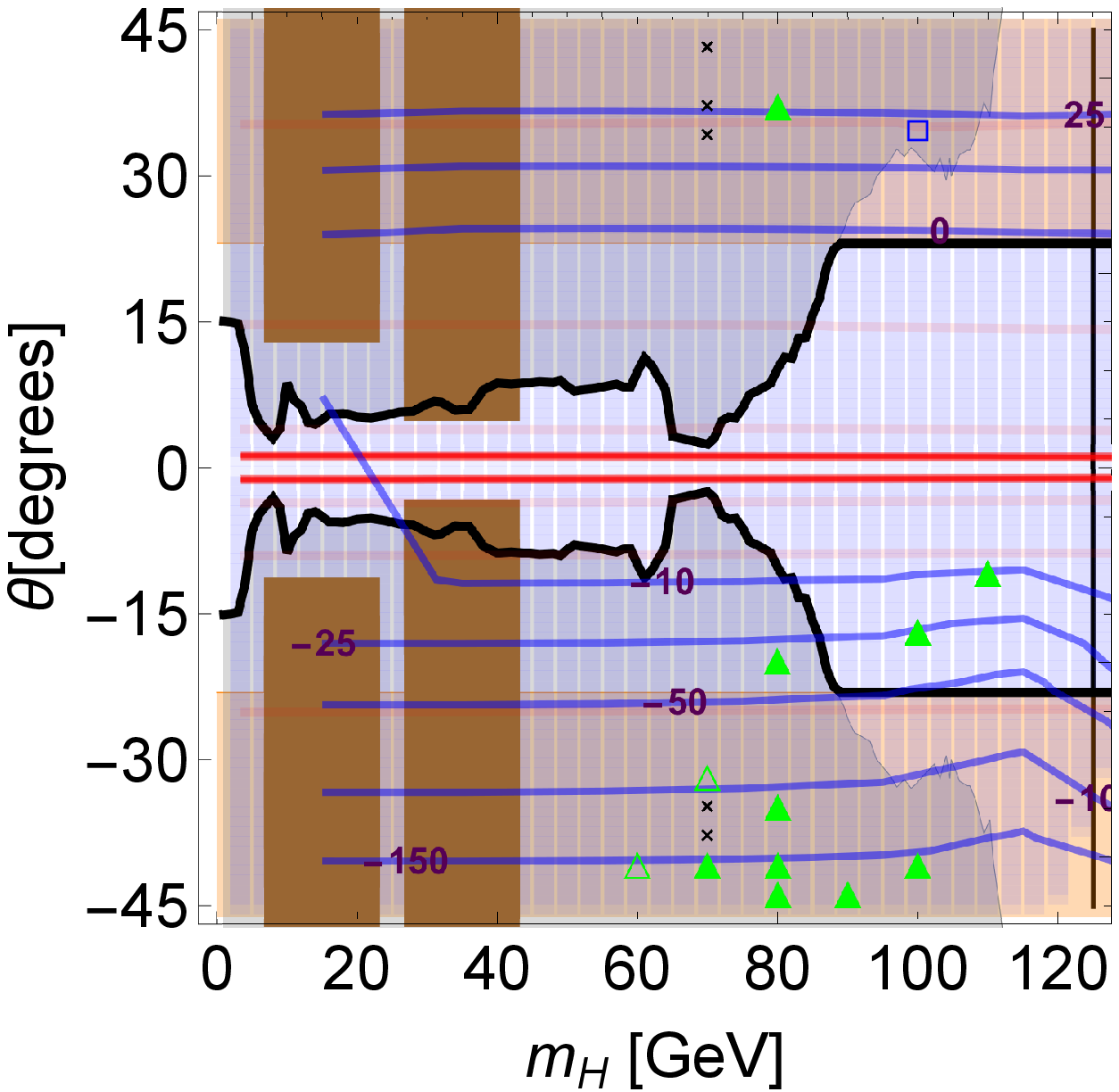}
\includegraphics[width=.49\textwidth]{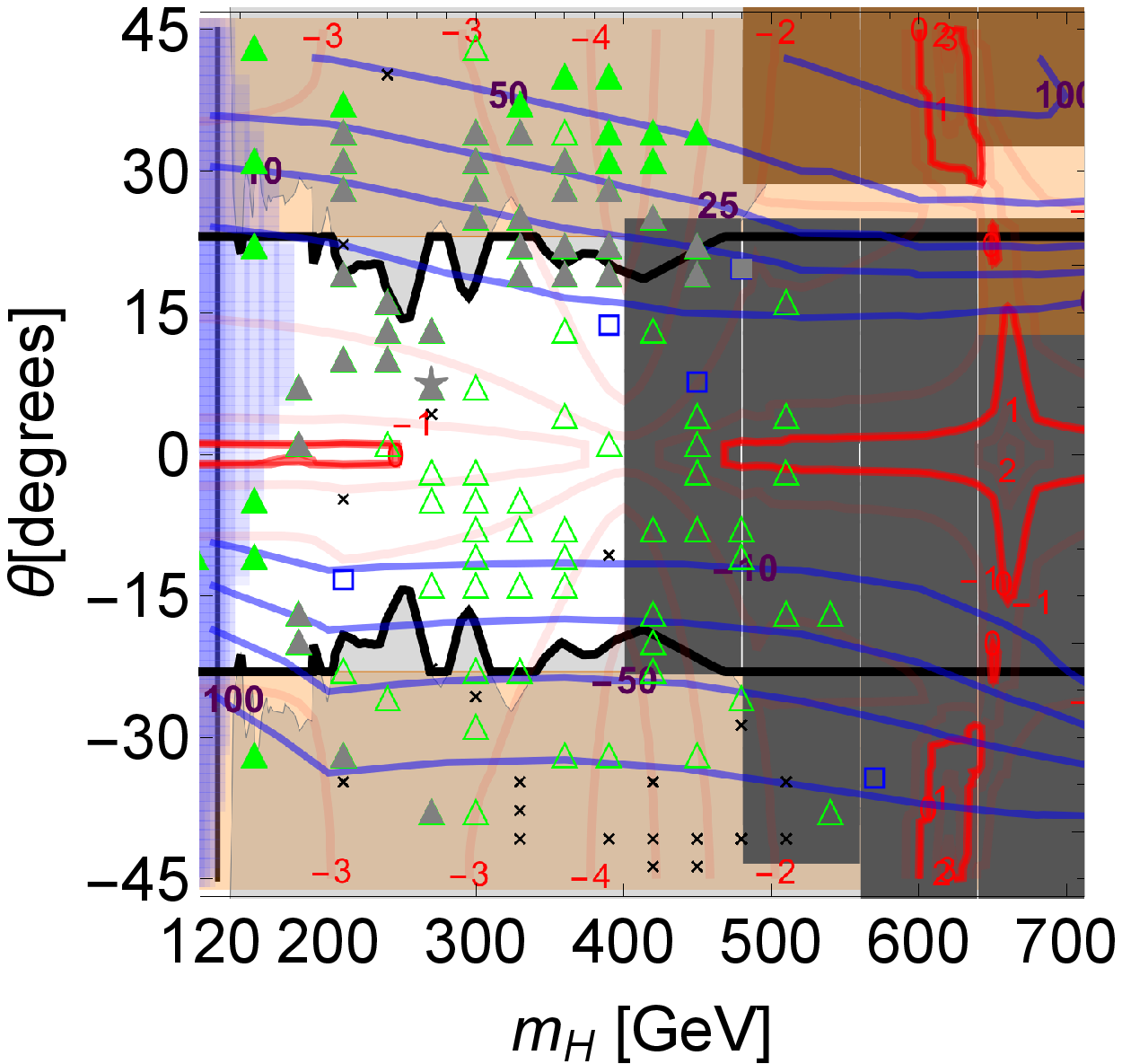}
\caption{
 Types of multi-step PT on the $(m_H^{},\theta)$ plane for the benchmark point $m_X=200~\GeV$ and $g_X=2$ ($v_S=100~\GeV$).
 The left (right) frame shows light heavy $H$ cases: $m_H^{}<m_h^{}$ (heavy $H$ cases: $m_H^{}>m_h^{}$).
 Parameter sets predicting one-step PT with 1st order are marked with blue closed square, one-step PT with 2nd order with blue open square, two-step PT where both transitions are 1st order with green closed star, two-step PT where the latter one is 1st order with green closed triangle, and two-step PT where the former one is 1st order with green open triangle.
 The blue solid lines show the contours of $\Delta \lambda_{hhh}$ in percentage.
 The black lines show the combined exclusion limit obtained by $\kappa_Z^{}$ measurement (orange) and direct searches for $H$ (gray).
 In Model B, the following constraints on DM properties should be imposed.
 The red lines show the contours of the normalized relic density $\Omega_X^{}/\Omega_{\rm obs}^{}$ in common logarithm (The red bold lines show the cases of $\Omega_X^{}=\Omega_{\rm obs}^{}$).
 The cyan regions are excluded by DM direct detection by XENON1T in terms of $\log_{10}^{}(\sigma_X^{} \times \Omega_X^{}/\Omega_{\rm obs}^{}$).
 For more details, see TABLE~\ref{table:legends}.
}
\label{fig:200+1_1}
\end{figure}

\begin{figure}[t]
\centering
\includegraphics[width=.49\textwidth]{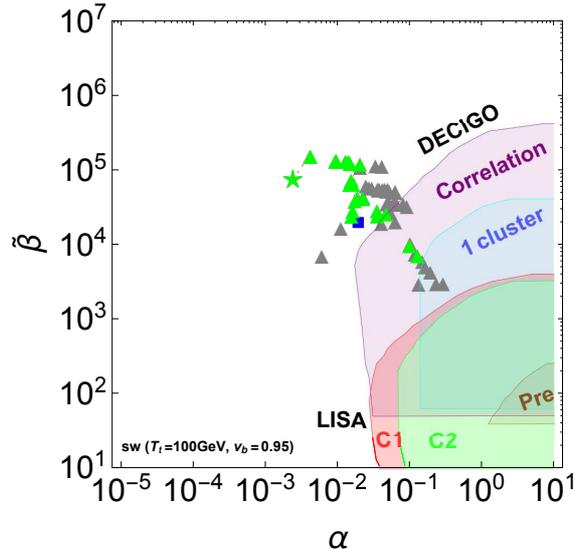}
\caption{
 Predicted values of $\alpha$ and $\widetilde{\beta}$ for the benchmark point $m_X^{}=200~\GeV$ and $g_X^{}=2$ ($v_S^{}=100~\GeV$).
 For two-step PT with 1st order $\to$ 1st order, the former (green closed star) and the latter (green closed triangle) is  connected with a red dashed line.
 Parameter sets excluded by the collider bounds shown in Fig.~\ref{fig:200+1_1} are marked with darker points.
 The expected sensitivities of LISA and DECIGO detector designs are computed by using the sound wave contribution for $T_t^{}=100~\GeV$ and $v_b^{}=0.95$.
}
\label{fig:200+1_2}
\end{figure}

\begin{figure}[t]
\centering
\includegraphics[width=.32\textwidth]{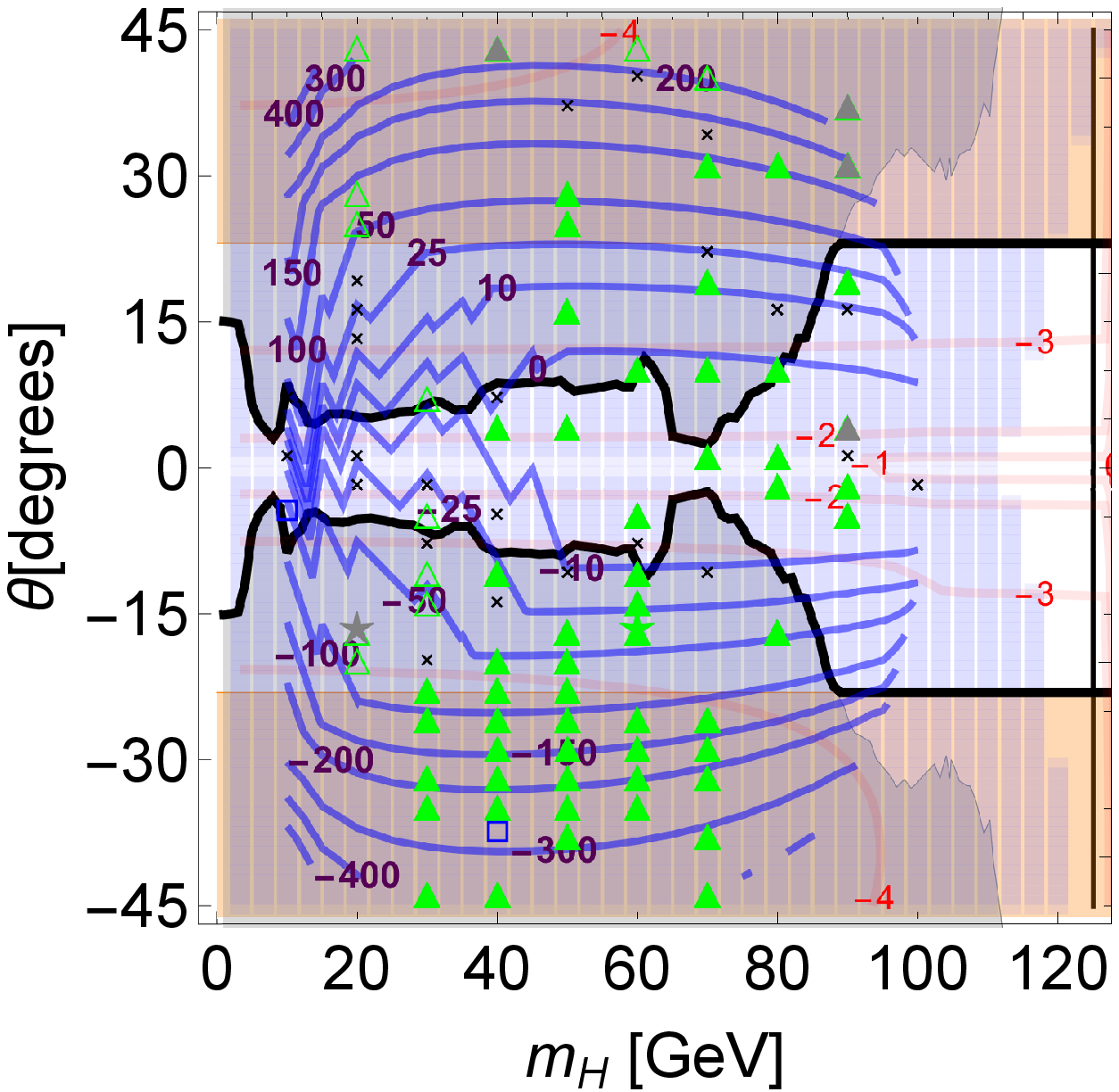}
\includegraphics[width=.32\textwidth]{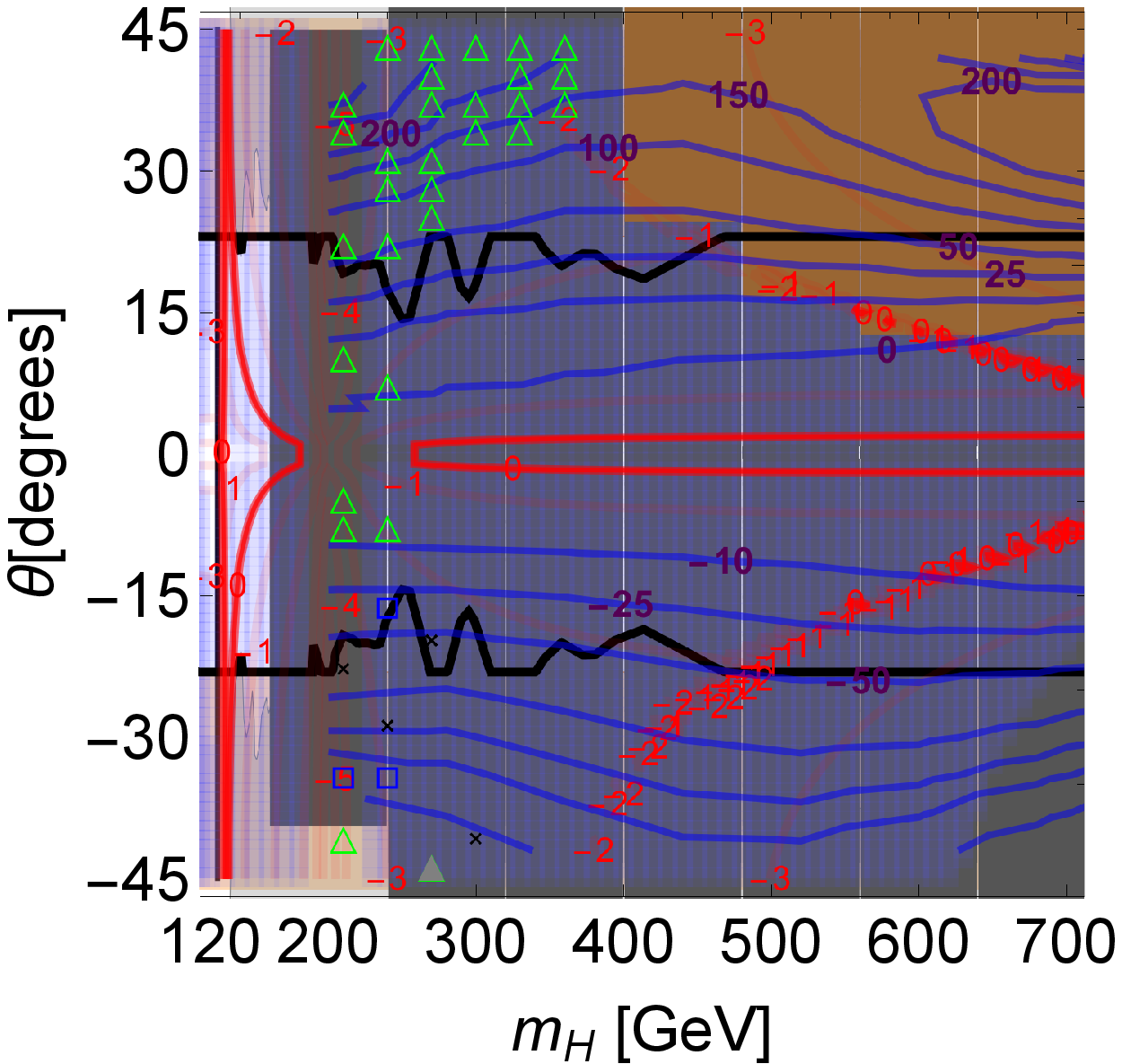}
\includegraphics[width=.32\textwidth]{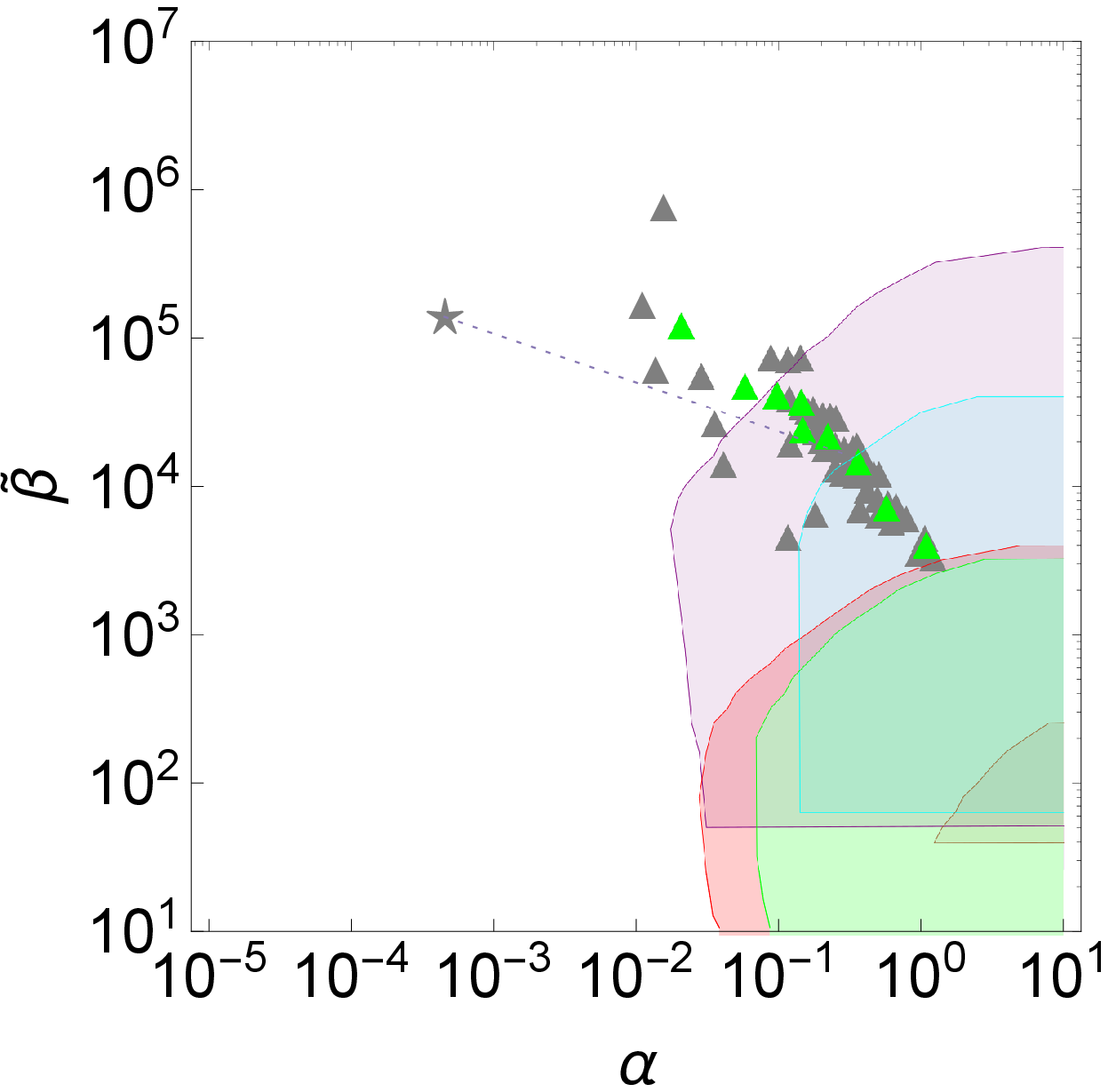}
\vspace{-3mm}
\caption{
 Types of multi-step PT for $m_X=100~\GeV$, $g_X=2$ ($v_S=50~\GeV$).
 See the captions of Figs.~\ref{fig:200+1_1}--\ref{fig:200+1_2}.
 }
 \label{fig:100+1}
\vspace{6mm}
\includegraphics[width=.32\textwidth]{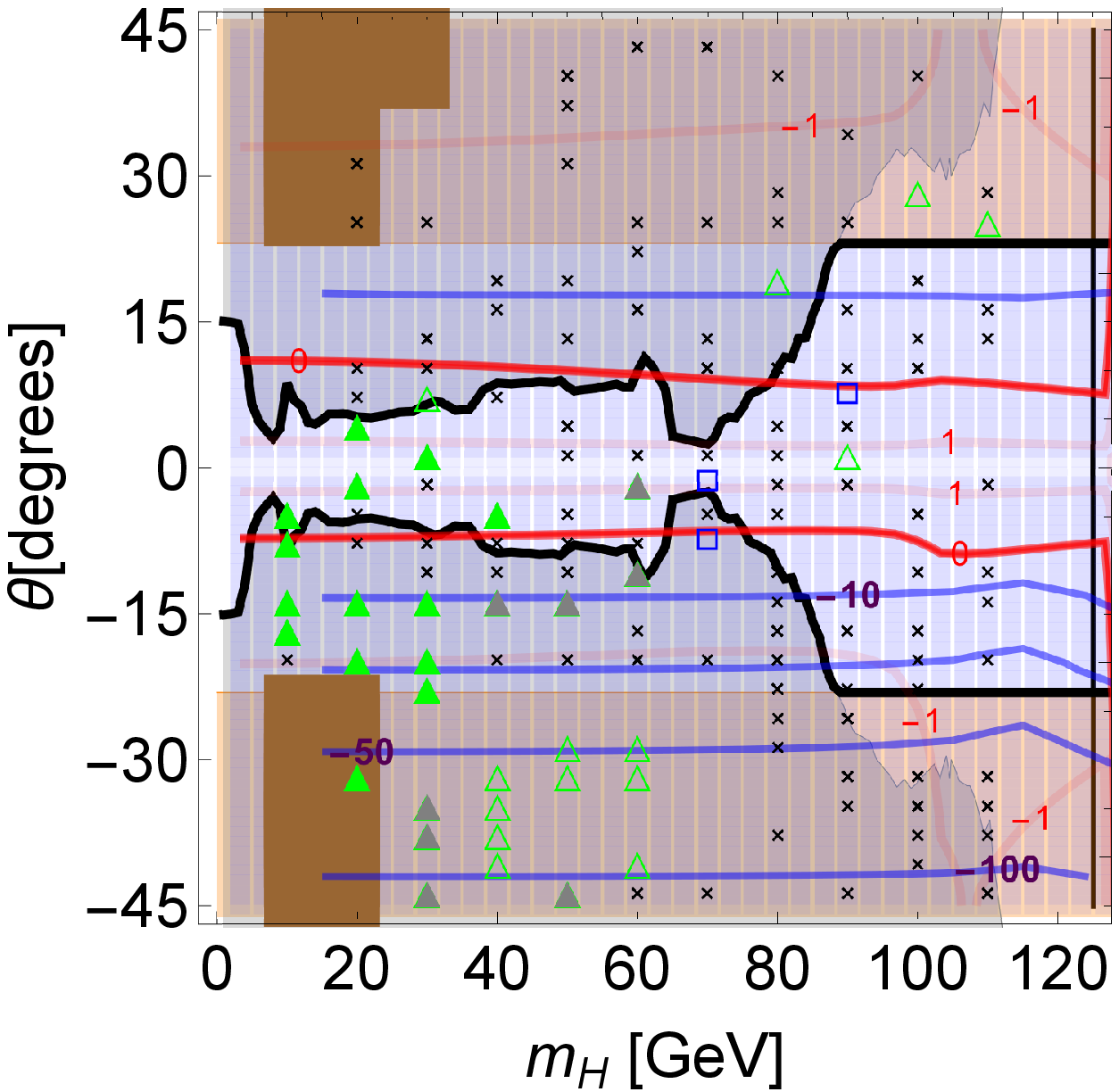}
\includegraphics[width=.32\textwidth]{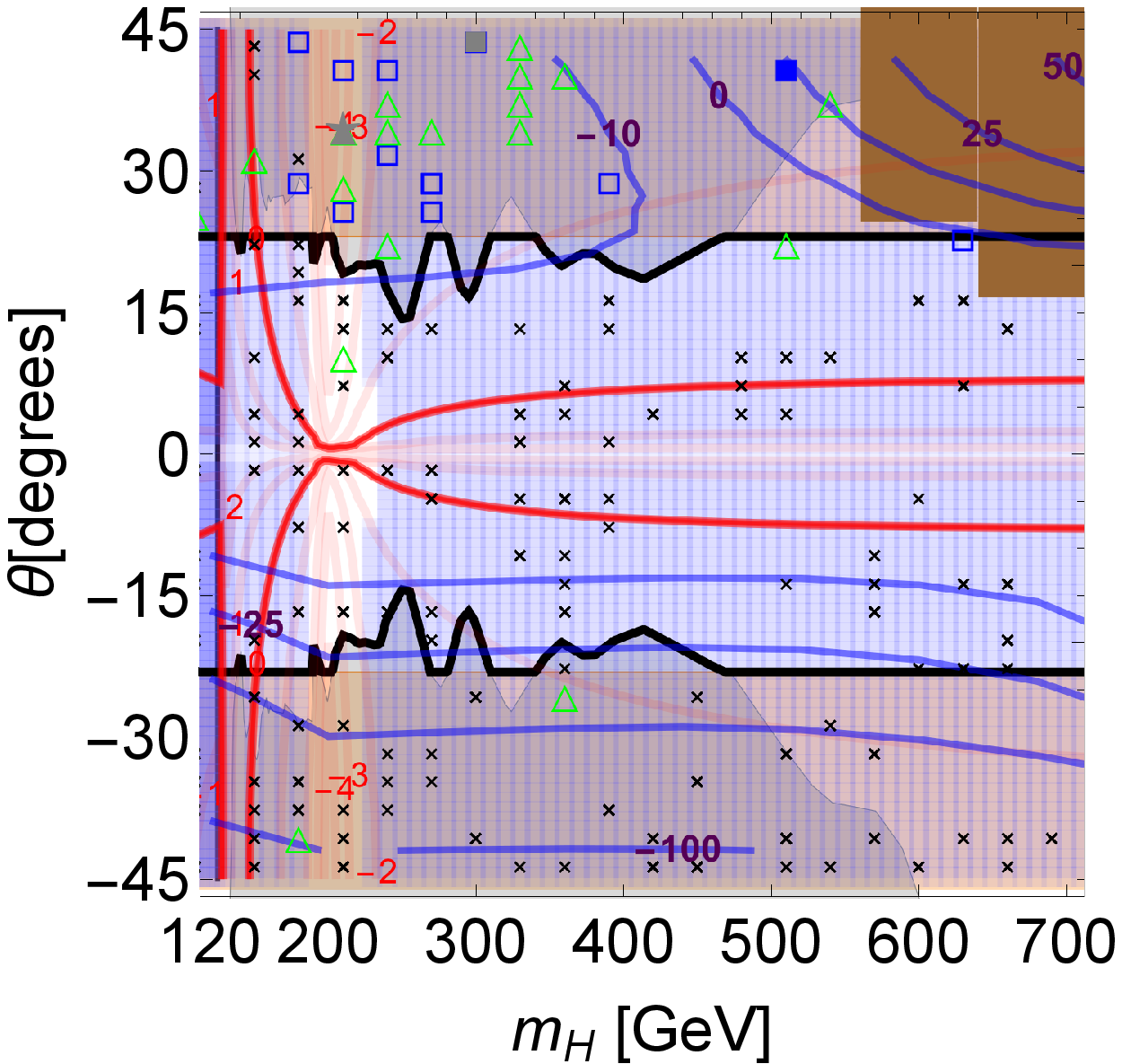}
\includegraphics[width=.32\textwidth]{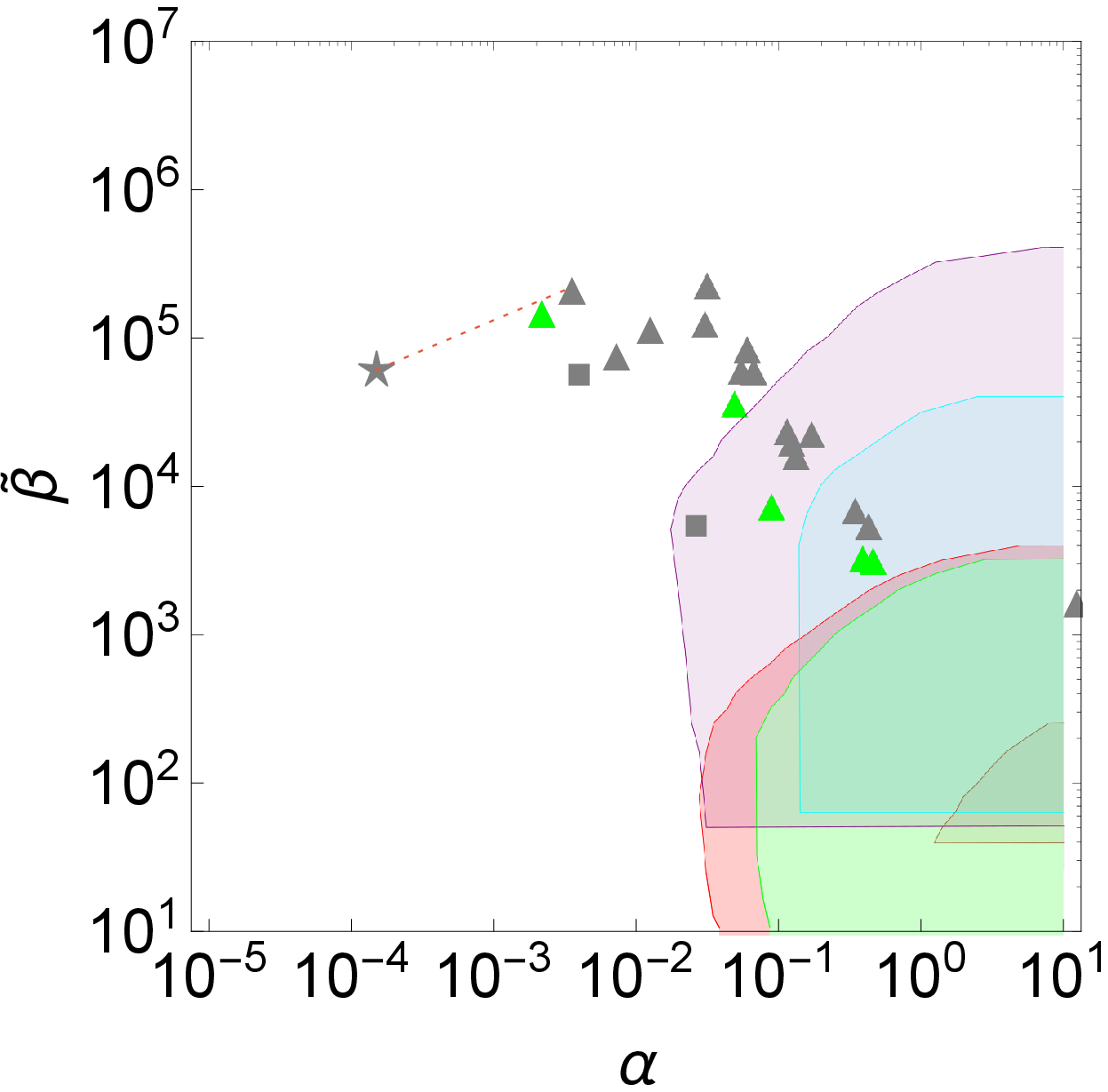}
\vspace{-3mm}
\caption{
 Types of multi-step PT for $m_X=100~\GeV$, $g_X=0.5$ ($v_S=200~\GeV$).
 See the captions of Figs.~\ref{fig:200+1_1}--\ref{fig:200+1_2}.
 }
 \label{fig:100-1}
\end{figure}

\begin{figure}[t]
\centering
\includegraphics[width=.32\textwidth]{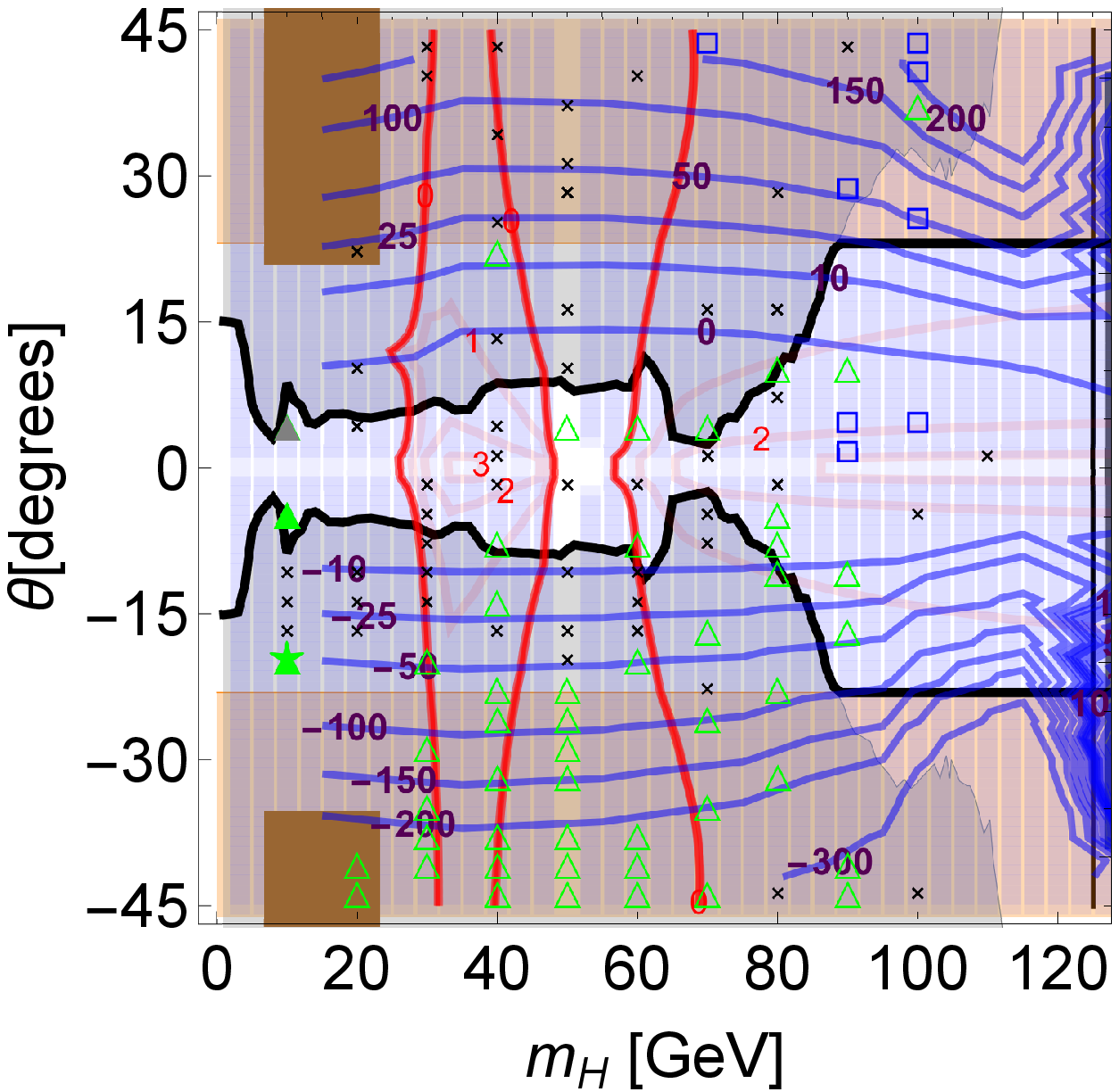}
\includegraphics[width=.32\textwidth]{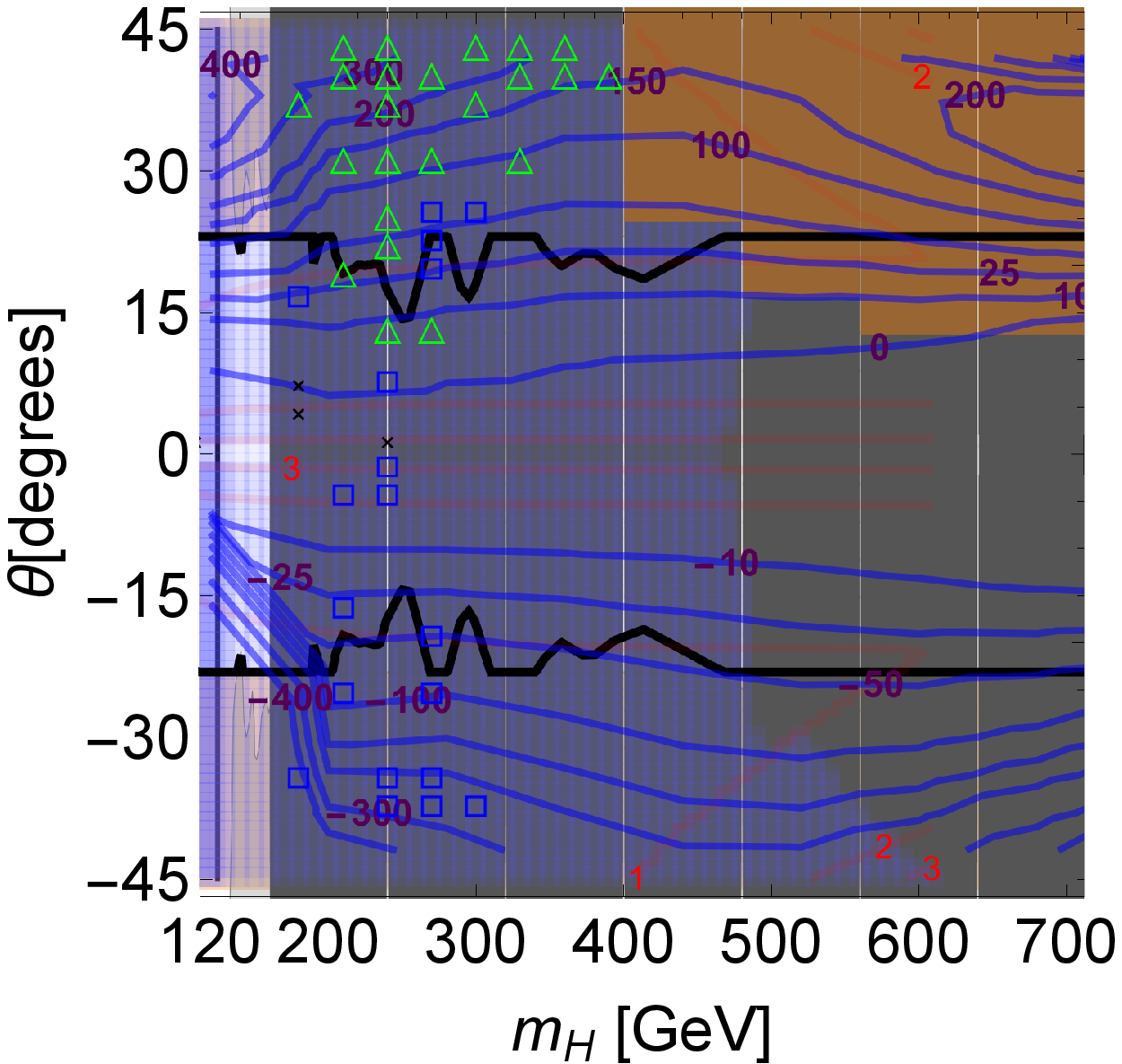}
\includegraphics[width=.32\textwidth]{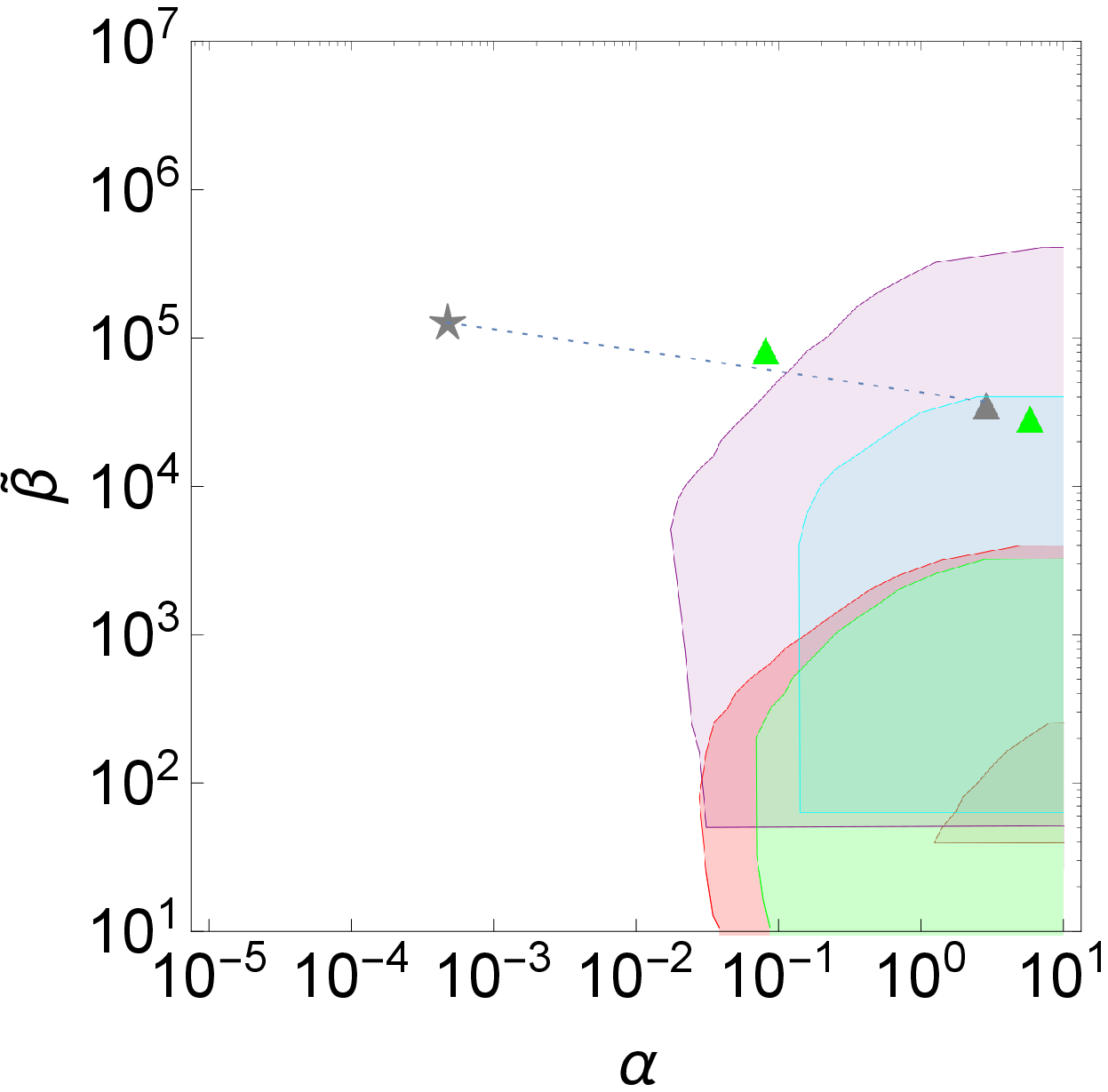}
\vspace{-3mm}
 \caption{
 Types of multi-step PT for $m_X=25~\GeV$, $g_X=0.5$ ($v_S=50~\GeV$).
 See the captions of Figs.~\ref{fig:200+1_1}--\ref{fig:200+1_2}.
}
 \label{fig:25-1}
\vspace{6mm}
\includegraphics[width=.32\textwidth]{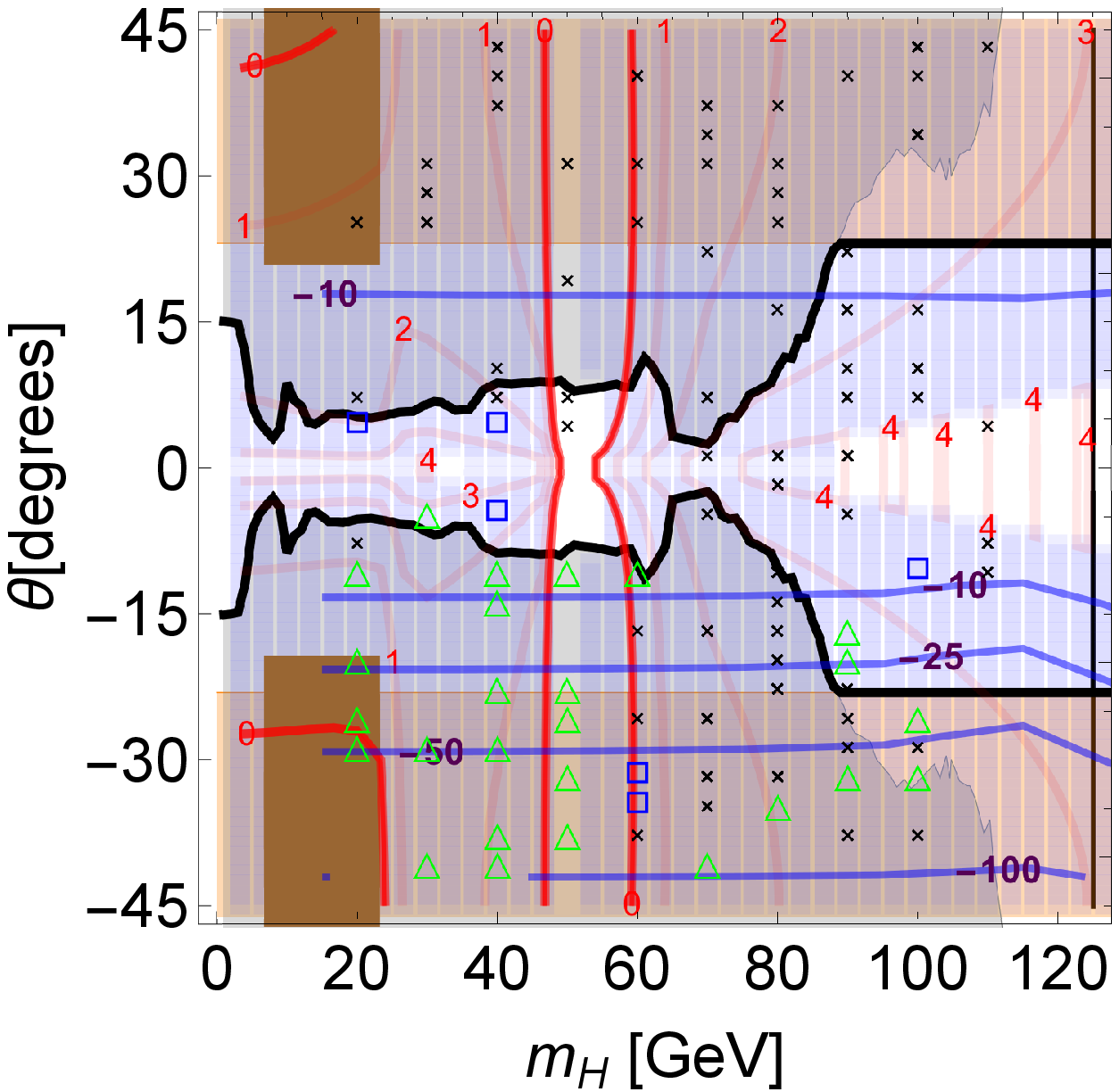}
\includegraphics[width=.32\textwidth]{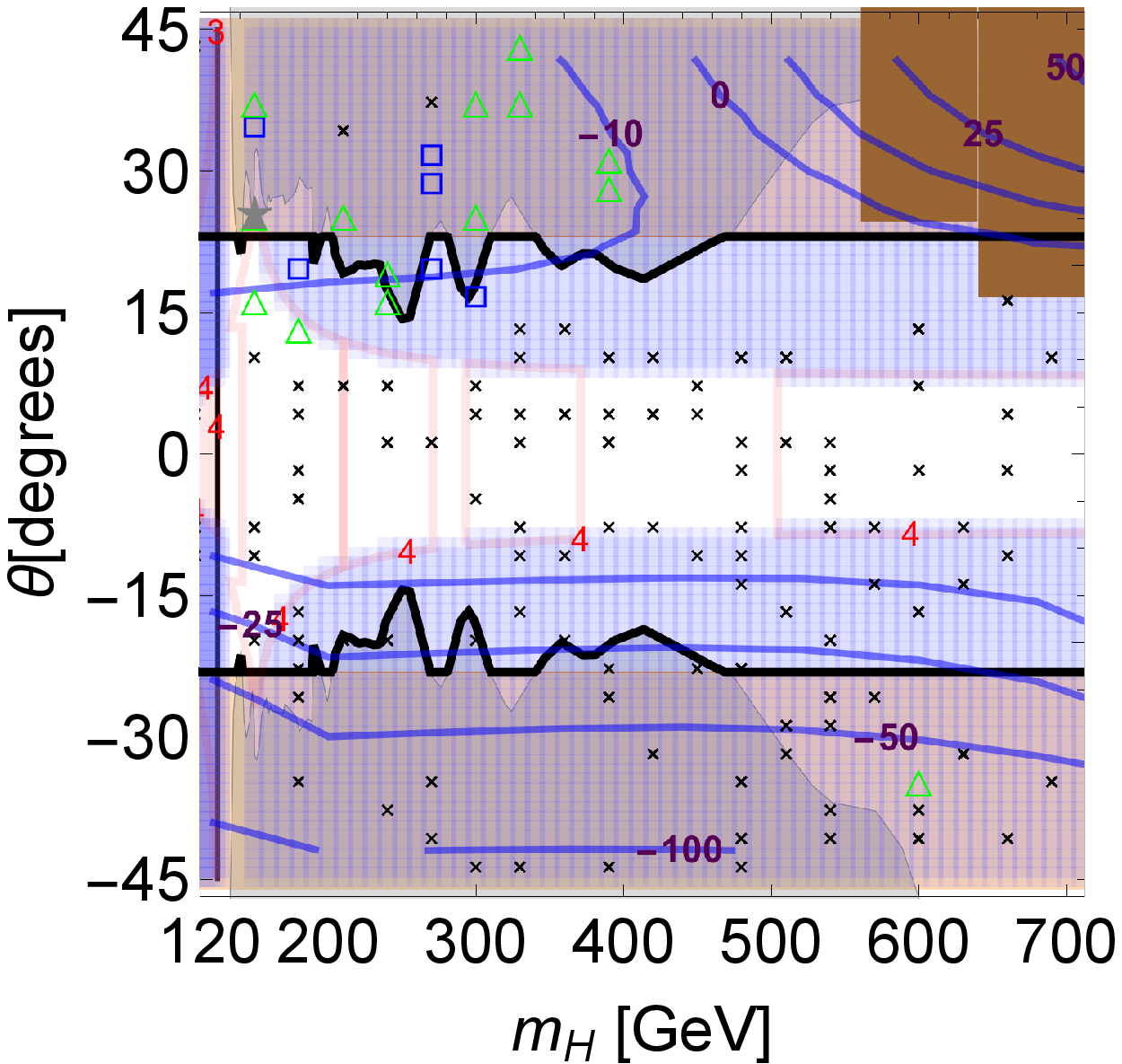}
\includegraphics[width=.32\textwidth]{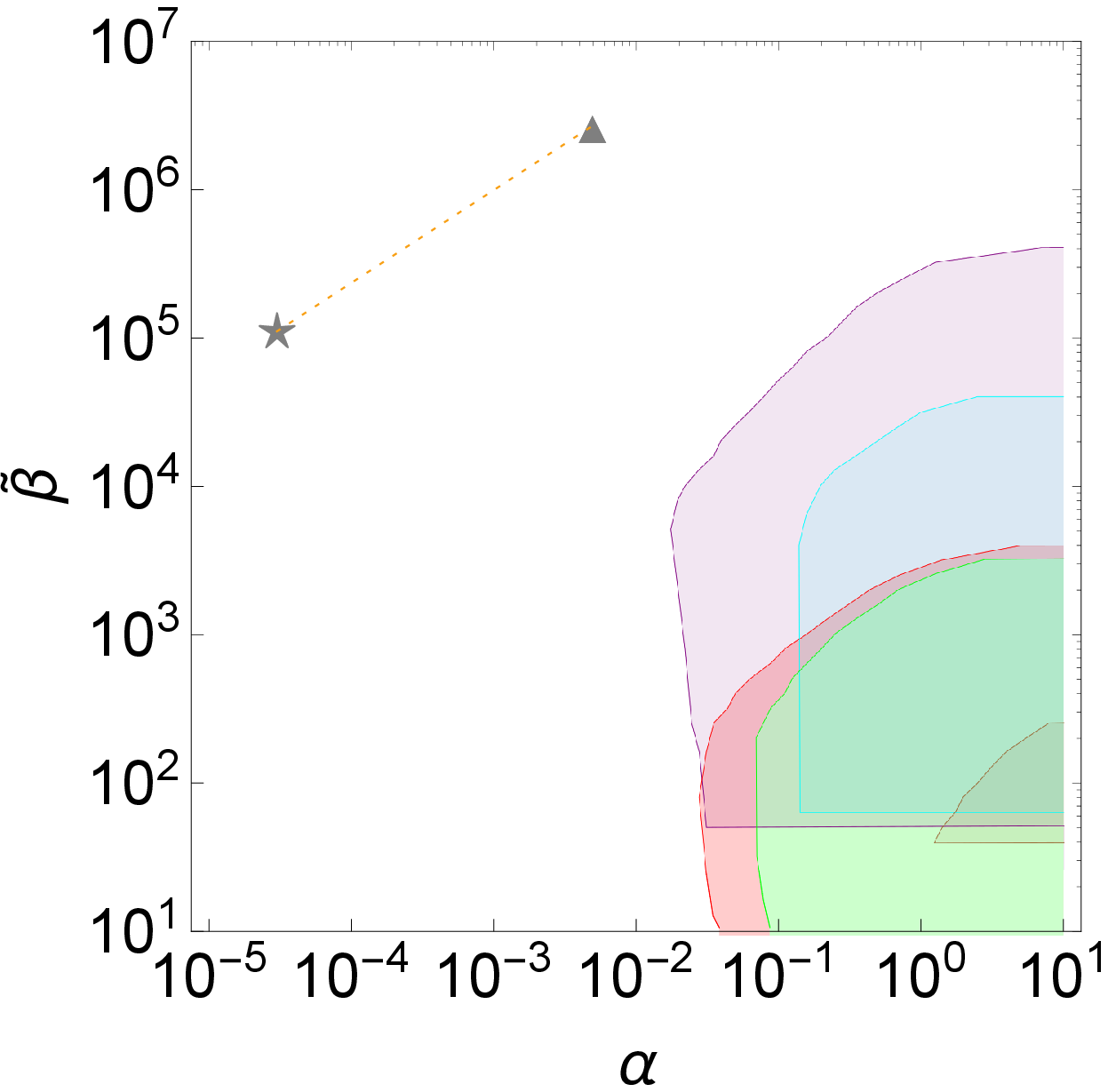}
\vspace{-3mm}
\caption{
 Types of multi-step PT for $m_X=25~\GeV$, $g_X=0.125$ ($v_S=200~\GeV$).
 See the captions of Figs.~\ref{fig:200+1_1}--\ref{fig:200+1_2}.
}
\label{fig:25-3}
\end{figure}

\begin{figure}[t]
\centering
\vspace{6mm}
\includegraphics[width=.32\textwidth]{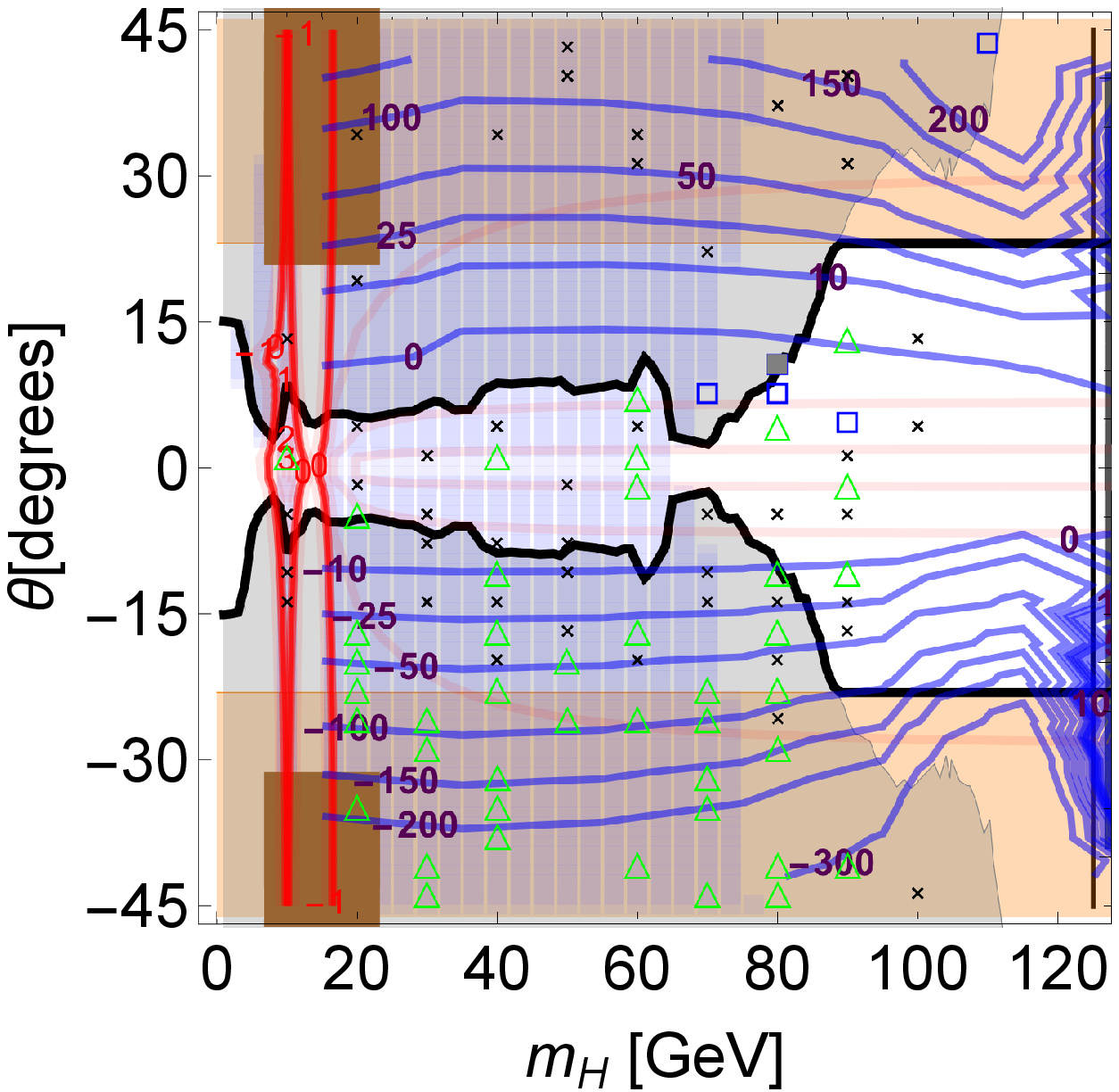}
\includegraphics[width=.32\textwidth]{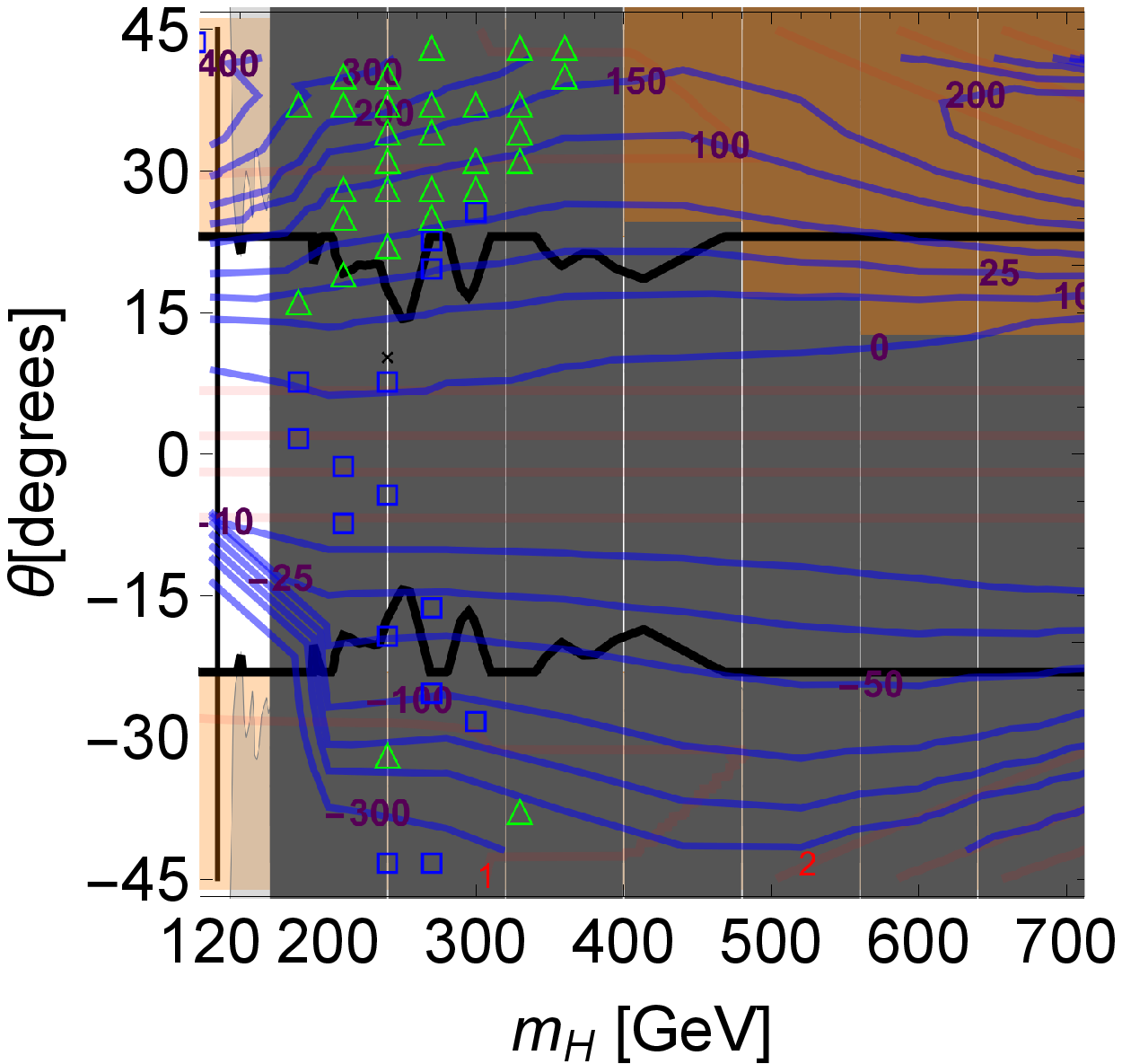}
\includegraphics[width=.32\textwidth]{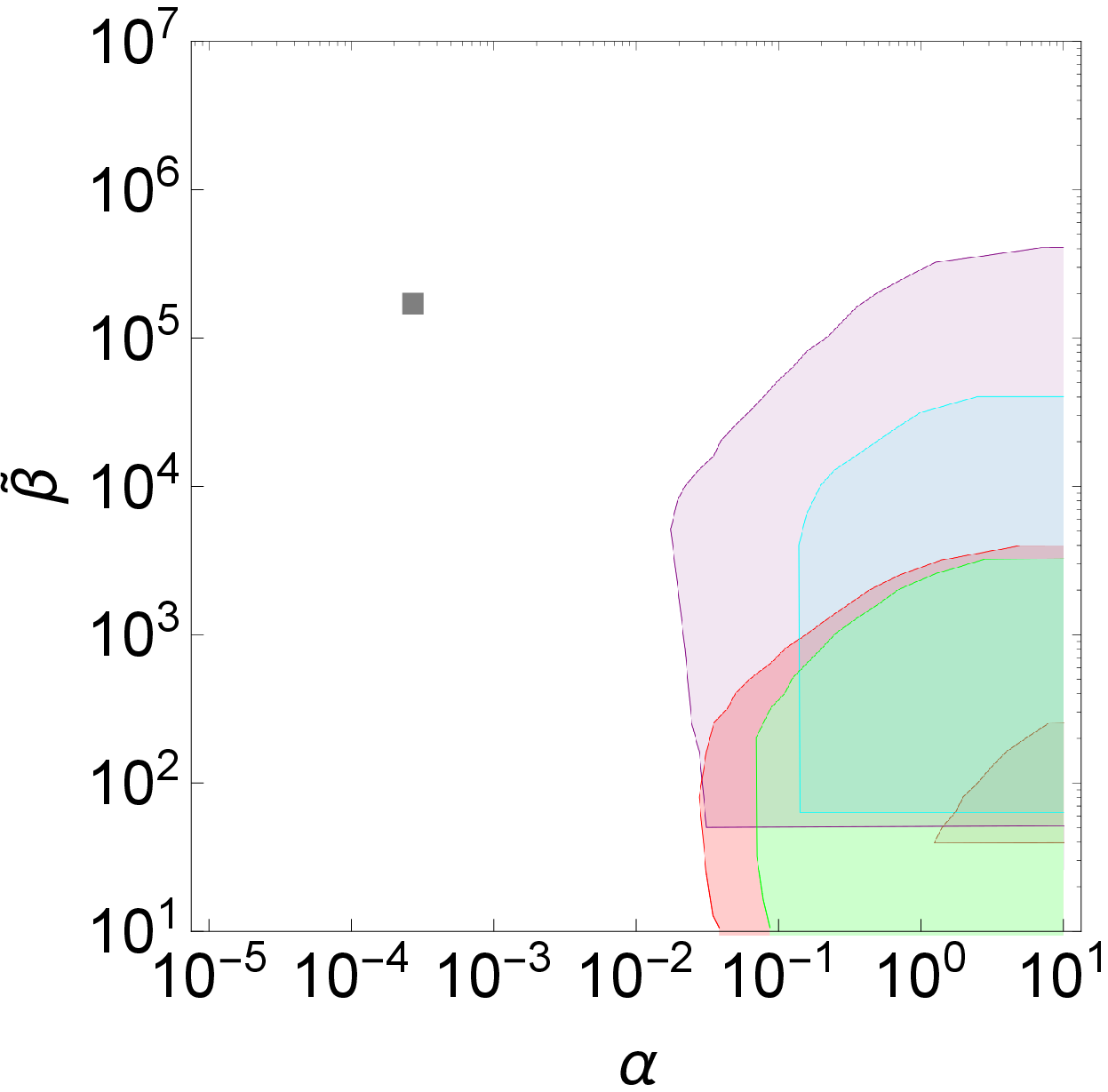}
\vspace{-3mm}
\caption{
 Types of multi-step PT for $m_X=6.25~\GeV$, $g_X=0.125$ ($v_S=50~\GeV$).
 See the captions of Figs.~\ref{fig:200+1_1}--\ref{fig:200+1_2}.
}
 \label{fig:6-3}
\end{figure}

 Our numerical results about the EWPT and GW signals for the six benchmark points defined in Fig.~\ref{fig:benchmark} are shown in Figs.~\ref{fig:200+1_1}, \ref{fig:200+1_2}, \ref{fig:100+1}, \ref{fig:100-1}, \ref{fig:25-1}, \ref{fig:25-3} and \ref{fig:6-3}.   
 The figure legends used for Figs.~\ref{fig:200+1_1}--\ref{fig:6-3} are listed in Table~\ref{table:legends}.

 In the left (right) frame of Fig.~\ref{fig:200+1_1} as well as in the left (middle) frames of Figs.~\ref{fig:100+1}--\ref{fig:6-3}, parameter sets that predict first order EWPT are marked with closed symbols for light $H$ cases: $m_H^{} < m_h^{}$ (heavy $H$ cases: $m_h < m_H$).
 There are five different types of PT: One-step PT with 1st order (blue closed square), one-step PT with 2nd order (blue open square), two-step PT where both transitions are 1st order (green closed star), two-step PT where the latter one is 1st order (green closed triangle), and two-step PT where the former one is 1st order (green open triangle)~\footnote{
 There are possibilities that crossover occurs at some parameter sets instead of 2nd order phase transition. 
 Since both the cases do not contribute to GWs, we collectively call them "2nd order" in this paper.
}.
 The blue solid lines show the contours of $\Delta \lambda_{hhh}$ in percentage.  
 The colored regions are excluded by perturbative unitarity (black), vacuum stability (brown), $\kappa_Z^{}$ measurement (orange), direct $H$ searches (gray), 
 In Model B, the massive $U(1)_X^{}$ gauge boson is stable and becomes a candidate for DM.
 Constraints on DM properties should be applied to the model parameter space.
 The red solid lines show the contours of the normalized relic DM abundance $\Omega_X/\Omega_{\rm obs}$ in common logarithm.
 The cyan regions are excluded by DM direct detection by XENON1T~\cite{Aprile:2017iyp}).

 The expected accuracy of the measurements of the Higgs boson couplings are as follows. 
 The high-luminosity LHC with $\sqrt{s}=14$~TeV and $L=3000~\ifb$ can constrain $\Delta \kappa_V$ with an accuracy of $2\%$~\cite{CMS:2013xfa} and 
$-0.8$ $(-0.2)$ $\lesssim \Delta \lambda_{hhh}/\lambda_{hhh}^\SM \lesssim$ 7.7 (2.6)~\cite{LHChhh2} (with a full kinematic analysis~\cite{Kling:2016lay}).
 Future electron-positron colliders can considerably ameliorate the precision.
 The stage of the ILC with $\sqrt{s}=250$~GeV and $L=2000~\ifb$ can limit $\Delta \kappa_W^{}$ to 1.8\% and $\Delta \kappa_Z^{}$ to 0.38\%~\cite{Fujii:2017vwa}.
 If the ILC with $\sqrt{s}=1~{\rm TeV}$ is realized, the $hhh$ coupling can be measured with an accuracy of $16\%$ ($10\%$) for $L=2000~\ifb$ ($5000~\ifb$)~\cite{Fujii:2015jha}. 
 The limit obtained from direct searches for the $H$-boson is discussed in Ref.~\cite{Chang:2017ynj} for the small mass region and in Ref.~\cite{Carena:2018vpt} for the large mass region.
 The high-luminosity LHC will extend the discovery reach.
 Therefore, the measurements of the properties of the Higgs bosons play an important role in pinning down viable model parameters.

 The detectability of GWs from strongly 1stOPT is shown in Fig.~\ref{fig:200+1_2} and the right frames of Figs.~\ref{fig:100+1}--\ref{fig:6-3}.
 All the points of multi-step PT with the first order EWPT (closed blue square, closed green star and closed green triangle in Fig.~\ref{fig:200+1_1} and the left two frames of Figs.~\ref{fig:100+1}--\ref{fig:6-3}) are displayed in these right frames.
 Among these parameter sets, points surviving the collider bounds (the black solid line in the corresponding left and middle frames) are marked with colored symbols in these right frames while the gray points are not compatible with the collider bounds. 
 
 In these figures, the areas with light colors are within the expected reach of the future space-based interferometers, LISA~\cite{Klein:2015hvg,Caprini:2015zlo,PetiteauDataSheet} and DECIGO~\cite{Kawamura:2011zz}. 
 The expected sensitivities of different LISA (DECIGO) designs are labeled by ``C1'' and ``C2'' (``Correlation'',  ``1 cluster''  and ``Pre'') following Ref.~\cite{Caprini:2015zlo} (Ref.~\cite{Kawamura:2011zz}).
 Notice that the transition temperature $T_t$ depends on the model parameters, we take $T_t^{}=100~\GeV$ for the purpose of illustration.
 The amplitude of produced GWs is enhanced as the velocity of the bubble wall $v_b^{}$ is increased.
 Since the uncertainty of the evaluation of $v_b^{}$ is large, we here consider an optimistic case of $v_b=0.95$. 
 For successful EWBG scenarios, on the other hand, subsonic wall velocities are preferable.
 In such cases, detecting GWs require smaller $\widetilde{\beta}$ and larger $\alpha$.

\begin{table}[p]
\centering
\caption{
Figure legends for Figs.~\ref{fig:200+1_1}--\ref{fig:6-3}.
}
\label{table:legends}
\vspace{3mm}
   \begin{tabular}{|c|c||l|} \hline
Categories & Symbols & Legends \\ \hline \hline
Theory	& $\colorbox{black!80}{\color{black!80} [\qquad]}$ & excluded by perturbative unitarity \\ \cline{2-3}
		& $\colorbox{brown}{\color{brown} [\qquad]}$ & excluded by vacuum stability \\ \hline
Collider	& {\color{blue} \rule{15mm}{0.4mm}} & contours of $\Delta \lambda_{hhh}/\lambda_{hhh}^\SM$ (\%) \\ \cline{2-3}
	 	& $\colorbox{orange!40}{\color{orange!40} [\qquad]}$ & constraint by the $\kappa_Z^{}$ measurement (see Eq. (\ref{eq:Hdirect})) \\ \cline{2-3}
		&  $\colorbox{gray!40}{\color{gray!40} [\qquad]}$ & constraint by the direct searches for the $H$-boson (see Ref.~\cite{Robens:2015gla}) \\ \cline{2-3}
		& {\color{black} \rule{15mm}{0.8mm}} & combined exclusion limit ($\colorbox{orange!40}{\color{orange!40} [\qquad]}$+$\colorbox{gray!40}{\color{gray!40} [\qquad]}$) \\ \hline
PT	& ${\color{blue} \blacksquare}$	& one-step PT (1st order) \\ \cline{2-3}
	& ${\color{blue} \square}$		& one-step PT (2nd order) \\ \cline{2-3}
	& ${\color{green} \bigstar}$		& two-step PT (1st order $\to$ 1st order) \\ \cline{2-3}
	& ${\color{green} \blacktriangle}$	& two-step PT (2nd order $\to$ 1st order) \\ \cline{2-3}
	& ${\color{green} \triangle}$		& two-step PT (1st order $\to$ 2nd order) \\ \cline{2-3}
	& (${\color{gray} \blacksquare}$, ${\color{gray} \bigstar}$, ${\color{gray} \blacktriangle}$) in ($m_H$, $\theta$)		& insensitive at GW observation  \\ \cline{2-3}
	& (${\color{gray} \blacksquare}$, ${\color{gray} \bigstar}$, ${\color{gray} \blacktriangle}$) in ($\alpha$, $\widetilde{\beta}$)		& excluded by the collider constraints ({\color{black} \rule{15mm}{0.8mm}}) \\ \hline
DM	& {\color{red} \rule{15mm}{0.8mm}} & contours of log$_{10}(\Omega_X/\Omega_{\rm obs})$ \\ \cline{2-3}
	& $\colorbox{blue!40}{\color{blue!40} [\qquad]}$ & excluded by XENON1T~\cite{Aprile:2017iyp}
	in terms of $\log_{10}(\sigma_X \times \Omega_X/\Omega_{\rm obs}$) \\ \hline
GW	& $\colorbox{purple!40}{\color{purple!40} [\qquad]}$ & DECIGO (Correlation) \\ \cline{2-3}
	& $\colorbox{cyan!40}{\color{cyan!40} [\qquad]}$ & DECIGO (1 cluster) \\ \cline{2-3}
	& $\colorbox{brown!40}{\color{brown!40} [\qquad]}$ & DECIGO (Pre) \\ \cline{2-3}
	& $\colorbox{red!40}{\color{red!40} [\qquad]}$ & LISA (C1) \\ \cline{2-3}
	& $\colorbox{green!40}{\color{green!40} [\qquad]}$ & LISA (C2) \\ \hline
   \end{tabular}
\vspace{6mm}
\caption{
 Summary of our numerical results.
 For each benchmark points, the types of the multi-step PT are listed and the different solutions in terms of the detectability of the GW observations and the DM constraints are labeled.
 In the GW column, $\bigcirc$ ($\triangle$) denotes that there are (no) regions which reach to the planned sensitivity of LISA and DECIGO.
 In the DM column for Model B, $\bigcirc$ ($\triangle$) denotes that the VDM and nucleon elastic scattering cross section rescaled with relic density satisfy the bound from XENON1T~\cite{Aprile:2017iyp} with the predicted relic density which accommodates the observed one (a subdominant DM component).
}
\label{table:summary}
\vspace{3mm}
   \begin{tabular}{|c|c||c|c|c|c||c|c|} \hline
$(m_X[\GeV], g_X)_{v_s}$ & Fig. & $\#$ & PT & GW & DM & Refs. \\ \hline \hline
$(200, 2)_{100\GeV}$ & \ref{fig:200+1_1}, \ref{fig:200+1_2} & 1 & C & $\bigcirc$ & excluded by XENON1T & \\ \cline{3-7}
                                          & & 2 & C & $\triangle$ & $\bigcirc$ & VDM~\cite{Baek:2012se,Duch:2015jta} \\ \cline{3-7}
                                          & & 3 & D & $\triangle$ & $\triangle$ & \\ \hline
$(100, 2)_{50\GeV}$ & \ref{fig:100+1} & 4 & C & $\bigcirc$ & excluded by XENON1T & \\ \hline
$(100, 0.5)_{200\GeV}$ & \ref{fig:100-1} & 5 & C & $\bigcirc$ & excluded by XENON1T                       & \\ \cline{3-7}
                                           & & 6 & D & excluded by collider & excluded by XENON1T & HSM~\cite{Fuyuto:2014yia} \\ \hline
$(25, 0.5)_{50\GeV}$ & \ref{fig:25-1} & 7 & C & $\bigcirc$ & excluded by XENON1T                      & \\ \hline
$(25, 0.125)_{200\GeV}$ & \ref{fig:25-3} & 8 & C & excluded by collider & excluded by XENON1T & \\ \hline
$(6.25, 0.125)_{50\GeV}$ & \ref{fig:6-3}& 9 & D & excluded by collider & excluded by XENON1T & \\ \hline
   \end{tabular}
\end{table}

\clearpage

 We summarize the results of our benchmark point study in Table~\ref{table:summary}.
 Several parameter sets in each benchmark point are classified in light of the detectability of the GWs
 at the future interferometers, the DM direct detection constraints by XENON1T.
 The patterns of PT are detailed as follows:
  \begin{itemize}
 \item {\bf One-step phase transition}: 
In the $U(1)_X$ gauge model, the limit of $g_X \to 0$ corresponds to the Higgs singlet model (HSM) with the spontaneously broken $Z_2$ symmetry, where a real isospin scalar singlet $S$ is introduced in addition to the Higgs doublet $\Phi$.
 In this case, the 1stOPT is realized as the one-step PT with type-D in Fig.~\ref{fig:PT}.
 As mentioned in Ref.~\cite{Fuyuto:2014yia}, however, such a case is excluded by the collider bounds.
 In contrast, we can find some points satisfying the collider bounds in the $U(1)_X$ model by the existence of the dark photon ($X^0$-boson) contribution.
 However, the strength of the PT is not so strong and it is not enough to detect by the future GW observations as shown by the blue point in Fig.~\ref{fig:200+1_2} and Figs.~\ref{fig:100+1}--\ref{fig:6-3}.
\item {\bf Two-step phase transition}:
 There are two cases for two-step PT with the first order EWPT with type-C in Fig.~\ref{fig:PT}: ``2nd order $\to$ 1st order'' or ``1st order $\to$ 1st order''.
 In the former case, EWPT with 1stOPT is shown by the triangle point in Figs.~\ref{fig:200+1_1}--\ref{fig:6-3}.
 In the latter case, two 1stOPTs can be calculated as shown by star and triangle point connected by the dashed line in Fig.~\ref{fig:200+1_2} and Figs.~\ref{fig:100+1}--\ref{fig:6-3}.   In most of the parameter region, the 1stOPT is strong and it can be detected by the future GW observations.
\end{itemize}

 As we can see from the results, large values of $m_X^{}$ ($\gtrsim 25~\GeV$) and large values of $g_X$ ($\gtrsim 0.5$) are preferred for detectable GW signals.
 We can understand analytically that $m_X$ is correlated with the detectability of GWs as follows.
 The difference of the vacuum energies of the I- and the EW phases as well as that of the II- and the EW phases are given by~\cite{Baek:2012se}
\begin{align}
\Delta V^{\rm (I)} \equiv V_0^{\rm (I)}(0, \bar{v}_S)-V_0^{\rm (EW)}(v_\Phi, v_S)&=\frac{v_\Phi^4}{16\lambda_S}(4\lambda_\Phi \lambda_S - \lambda_{\Phi S}^2)>0, 
\end{align}
and
\begin{align}
\Delta V^{\rm (II)} \equiv V_0^{\rm (II)}(\bar{v}_\Phi, 0)-V_0^{\rm (EW)}(v_\Phi, v_S)&=\frac{v_S^4}{16\lambda_\Phi}(4\lambda_\Phi \lambda_S - \lambda_{\Phi S}^2)>0, 
\end{align}
respectively, with
\begin{align}
\bar{v}_S=\pm\sqrt{v_S^2+\frac{\lambda_{\Phi S}v_\Phi^2}{2\lambda_S}}, 
\quad
\bar{v}_\Phi=\pm\sqrt{v_\Phi^2+\frac{\lambda_{\Phi S}v_S^2}{2\lambda_\Phi}}.
\end{align}
 Using Eq.~(\ref{eq:vs}), the EW vacuum becomes always the global minimum at the tree level.
 On the other hand, the latent heat is approximately given by the difference of the potential minima between the false vacuum and the true vacuum.
 As we know in our numerical results, since the detectable GWs are realized by the transitions of type-C in this model, the typical strength of the GWs is parametrized by 
\begin{align}
\alpha \propto \epsilon \simeq \Delta V^{\rm (II)} \propto v_S^4 \propto m_X^4.
\end{align}
 This correlation shows that $\alpha$ is controlled by $m_X$.
 In general, the strength of the GWs as the functions of $\alpha$ and $\widetilde{\beta}$, which is given in Ref.~\cite{Caprini:2015zlo}, is enhanced by increasing $\alpha$.
 In addition, we find that $g_X$ also contributes to the GW detectability as shown in  our numerical results, e.g. by comparing Fig.~\ref{fig:25-1} and Fig.~\ref{fig:25-3}.
 Notice that large values of $g_X$ (small values of $v_S$) also give the upper bound of $m_H$ by the perturbativity condition (see Appendix~\ref{sec:analytic}).
 As the result, the lower bound is at least $m_X^{}>100~\GeV$, but $m_X^{}>25~\GeV$ can be possible depending on $g_X$ contribution at narrow parameter space shown in the left frame of Fig.~\ref{fig:25-1}.
 DECIGO is capable of detecting stochastic GWs from the sound wave source in a part of the model parameter region with the strongly 1stOPT.
 It might be challenging to detect by LISA because all points are in the $\widetilde{\beta}>10^3$ region.

\section{Discussion and conclusions}
\label{sec:conclusions}

In the following, we summarize our results and discuss some relevant points one by one.

 In this paper, we define the Landau pole $\Lambda_{\LP}$ as the scale where any of the Higgs 
 couplings is as strong as~\cite{Hashino:2016rvx}
\begin{align}
 |\lambda_{\Phi,S,\Phi S}(\Lambda_{\LP})|=4\pi.
\label{eq:lp}
\end{align}
 The one-loop level $\beta$ functions for these couplings are provided in Appendix~\ref{sec:beta}.
 In the $U(1)_X^{}$ gauge model, the Landau pole can be above the Planck scale as discussed in Ref.~\cite{Duch:2015jta}.    
 Since our model has only one $\Phi$-$S$ mixing term, namely $|\Phi|^2 |S|^2$, we need large Higgs couplings for strongly first order EWPT to occur.
 Then, the Landau pole $\Lambda_{\LP}$ appears at around ${\cal O}(10^{4})~\GeV$~\footnote{
 Such a strongly-coupled extended Higgs sector may interpret as a low-energy effective theory of the new physics above $\Lambda_{\LP}$, see \cite{Kanemura:2012hr} for the supersymmetric gauge theory that cause confinement.
}.
 In contrast, if there are two $\Phi$-$S$ mixing terms, {\it e.g.} $|\Phi|^2 |S|^2$ and $|\Phi|^2 S$ in the Higgs singlet model, 
 the Landau pole can be as large as $\Lambda_{\LP} \sim {\cal O}(10^{14})~\GeV$~\cite{Fuyuto:2014yia}.

 In our $U(1)_X^{}$ gauge model, two-step PT along the path of the type C is realized in most of the parameter points predicting first order EWPT and detectable GWs, As discussed in Ref.~\cite{Funakubo:2005pu}, even if EWPT is strong enough to suppress the sphaleron process after the transition, the type C transition cannot produce a sufficient amount of baryon asymmetry.
 In view of this, we have not imposed the condition for strongly 1stOPT,
\begin{align}
\left. \frac{\varphi_\Phi}{T} \right|_{T_t^{}} \gtrsim \zeta_{\rm sph}(T_t^{}),
\label{eq:strongly1stOPT}
\end{align}
with $\zeta_{\rm sph}^{}$ being typically close to the unity.
 
 The wall velocity $v_b^{}$ is a key parameter describing the dynamics of the bubble wall.
 In generic, there is a tension between strong GWs and baryon asymmetry in EWBG scenarios. 
 The amplitude of GWs is suppressed by $v_b^p$ with $p\gtrsim 3$ for small wall velocity.
 Large wall velocity $v_b\sim 1$ is preferred for detecting GWs.  
 On the other hand, successful EWBG scenarios favor lower wall velocity $v_b^{}\lesssim 0.15-0.3$ (for the calculation of $v_b^{}$ in the singlet models, see Ref.~\cite{Kozaczuk:2015owa}), which allows the effective diffuse of particle asymmetries near the bubble wall front~\cite{diffuse}. 
 In Ref.~\cite{No:2011fi}, however, it is pointed out that EWBG is not necessarily impossible even in the case with large $v_b^{}$.
 Further discussion is beyond the scope of this paper.

 In Ref.~\cite{Addazi:2017gpt}, the complementarity of dark photon searches and GW observations is discussed for the mass region $m_X\simeq {\cal O}(10^{-2}-1)~\GeV$ in the cases with strongly 1stOPT in Model A.
 However, the collider constraints are not properly included. 
 Taking these collider bounds on the Higgs boson properties into consideration, we have found that GW signals are detectable only for larger dark photon mass, say $m_X^{}\gtrsim {\cal O} (25-100)~\GeV$.
 As shown in Ref.~\cite{He:2017zzr}, the recent data from LHCb~\cite{Aaij:2017rft} and LHC Run-II~\cite{Aaboud:2017buh} give constraints on $\epsilon$, 
which is roughly smaller than $10^{-2}$ at least, 
for the mass regions $10.6~\GeV<m_X<70~\GeV$ and $150~\GeV<m_X<350~\GeV$, respectively.
 It is also shown that the mass region $20~\GeV<m_X<330~\GeV$ can be constrained at future lepton colliders in Ref.~\cite{He:2017zzr}.
 We expect that 1stOPT with such a heavy dark photon will be tested by synergy between future GW observations and dark photon searches.

 As for Model B with VDM, constraints on DM properties are discussed in Appendix~\ref{sec:dm} and shown in Fig.~\ref{fig:200+1_1} and Figs.~\ref{fig:100+1}--\ref{fig:6-3}. 
 If the thermal relic density of the VDM fulfills the total amount of the observed DM, GWs signals cannot reach the future sensitivities of GW observations (solution 2 in Table~\ref{table:summary}).  
 On the other hand, in the case where only a fraction of the total DM is composed of VDM, there is an allowed region than can be probed by GW observations (solution 1 in Table~\ref{table:summary}).

 In conclusion, we have comprehensively explored models with a dark photon whose mass stems from spontaneous $U(1)_X^{}$ gauge symmetry breaking by the nonzero VEV of the dark Higgs boson $S$ in light of the patterns of PT and the detectability of GWs from strongly 1stOPT as well as various collider and theoretical bounds.
 After imposing these constraints on the model parameter space, we have found that GWs produced from multi-step PT can be detected at future observations such as LISA and DECIGO if the dark photon mass is $m_X^{} \gtrsim 25~\GeV$ with the $U(1)_X^{}$ gauge coupling being $g_X^{} \gtrsim 0.5$.
 Some of the parameter regions predicting detectable GWs are covered by the measurements of $\Delta \kappa_V^{}$ and $\Delta \lambda_{hhh}^{}$
 at future colliders including the HL-LHC and ILC.
 The model where the dark photon becomes a candidate for DM has been also investigated in view of the thermal relic density and the current constraint by DM direct detection.
 In order for the future interferometers to observe GW signals, the VDM component should be at most about 10\% of the total DM abundance. 
 Our results have been summarized in Table~\ref{table:summary} and Figs.~\ref{fig:200+1_1}--\ref{fig:6-3}.

\acknowledgments

 This work was supported, in part, by the Sasakawa Scientific Research Grant from The Japan Science Society (HK),
 Grant-in-Aid for Scientific Research on Innovative Areas, the Ministry of Education, Culture, Sports, Science and Technology, No.\ 16H01093 (MK),
 No.\ 17H05400 (MK)
 and No.\ 16H06492 (SK),
 Grant H2020-MSCA-RISE-2014~no.~645722 (Non Minimal Higgs) (SK),
 JSPS Joint Research Projects (Collaboration, Open Partnership) ``New frontier of neutrino mass generation mechanisms via Higgs physics at HC and flavor physics'' (SK),
 National Research Foundation of Korea (NRF) Research Grant NRF-2015R1A2A1A05001869 (PK,TM),
 and the NRF grant funded by the Korea government (MSIP) (No. 2009-0083526) through Korea Neutrino Research Center at Seoul National University (PK).

\appendix

\section{Perturbative unitarity}
\label{sec:analytic}

\begin{figure}[t]
\centering
\includegraphics[width=.49\textwidth]{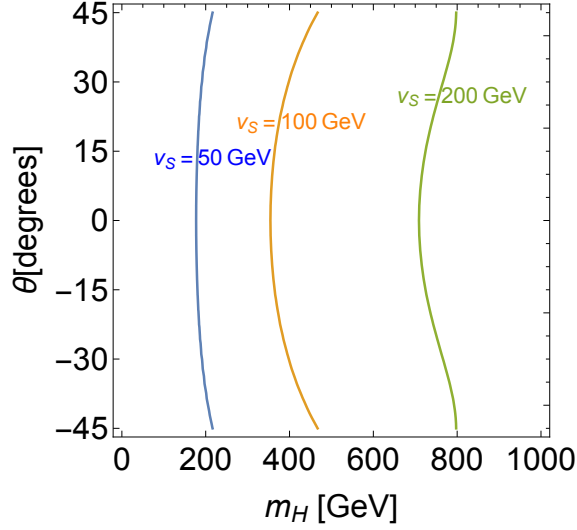}
\caption{
 Constraints from perturbative unitarity in the $(m_H^{}, \theta)$ plane for $v_S^{} = 50$~GeV (blue), 100~GeV (orange) and 200~GeV (green).
}
\label{fig:perturbativity}
\end{figure}
%
 Here, we discuss constrains from perturbative unitarity using analytic formulae at the tree level for illustrating the behavior of the model parameters. 
 The Higgs couplings in Eqs.~(\ref{eq:mass}) are expressed in terms of the masses of the Higgs bosons and the mixing angle as
\begin{align}
2 \lambda_\Phi v_\Phi^2 =m_{h}^2 c_\theta^2 +m_{H}^2 s_\theta^2, \quad
2 \lambda_S v_S^2 =m_{h}^2 s_\theta^2 +m_{H}^2 c_\theta^2, \quad
\lambda_{\Phi S} v_\Phi v_S =(m_{h}^2-m_{H}^2) c_\theta s_\theta. 
\end{align}
 Then, the constraints obtained from Eq.~(\ref{eq:perturbativity}) can be projected on the $(m_H^{}, \theta)$ plane as shown in Fig.~\ref{fig:perturbativity}.
 One can see that the excluded regions (indigo) in Fig.~\ref{fig:200+1_1} ($v_S=100~\GeV$) and Figs.~\ref{fig:100+1}, \ref{fig:25-1} and \ref{fig:6-3} ($v_S=50~\GeV$) are consistent with the corresponding contours in Fig.~\ref{fig:perturbativity}.

\section{One-loop renormalization group equations}
\label{sec:beta}

 The $\beta$ function for a coupling $\lambda$ is defined by $\beta (\lambda) \equiv d\,\lambda/d \log(\mu)$, with $\mu$ being the energy scale. 
 In our model, the one-loop renormalization group equations for the couplings in the Higgs sector and the gauge coupling $g_X^{}$ are given by~\cite{Baek:2012se,Duch:2015jta}
\begin{align}
\beta (\lambda_\Phi) &= \frac{1}{16\pi^2}
\left[
	24\lambda_\Phi^2+\lambda_{\Phi S}^2-6 y_t^4
	+\frac{3}{8}\Big\{2g_2^4+(g_2^2+g_1^2)^2\Big\}
	-\lambda_\Phi \Big\{3(3g_2^2+g_1^2)-12 y_t^2\Big\}
\right], \label{beta_lamH} \nonumber \\
\beta (\lambda_S) &= \frac{1}{16\pi^2} 
\Big[
	20\lambda_S^2+2\lambda_{\Phi S}^2+6g_X^4Q_S^4-12\lambda_S g_X^2Q_S^2
\Big], \nonumber \\
\beta (\lambda_{\Phi S}) &= \frac{1}{16\pi^2}
\left[
	\lambda_{\Phi S}(12\lambda_\Phi+8\lambda_S+4\lambda_{\Phi S})
	-\lambda_{\Phi S}\left\{\frac{3}{2}(3g_2^2+g_1^2)-6 y_t^2+6g_X^2Q_S^2 \right\}
\right], \nonumber \\
\beta (g_X) &= \frac{1}{16\pi^2}
	\frac{1}{3}g_X^3Q_S^3. \nonumber
\end{align}

\section{Dark matter relic abundance and direct detection}
\label{sec:dm}

 The parameter space of Model B is constrained in light of the thermal DM relic abundance and direct detection. 
 The observed DM abundance reported by the Planck collaboration is~\cite{Ade:2015xua}
\begin{align}
\Omega_{\rm obs}^{} h^2 = 0.1199\pm 0.0022.
\end{align}
 At a WIMP mass of $35~\GeV$, spin-independent DM-nucleon cross sections above
\begin{align}
\sigma^{\rm SI}_p=7.7\times 10^{-47}~{\rm  cm}^2
\end{align}
are excluded at the $90\%$ C.L. by XENON1T~\cite{Aprile:2017iyp}~\footnote{
 See also the recent results by LUX~\cite{Akerib:2016vxi} and PandaX-II~\cite{Cui:2017nnn}.
}.
 In Model B, the elastic cross section of $X$ scattering off the proton is obtained as~\cite{Baek:2012se}
\begin{align}
\sigma_X= \frac{c_\theta^2 s_\theta^2 \mu^2}{\pi}  \left[ {m_X^{} m_p^{} f_p^{} \over v v_s }  
\left( \frac{1}{ m_h^2 } -\frac{1 }{ m_H^2 }\right)\right]^2 ,
\label{eq:dm}
\end{align}
where $\mu_X^{} = m_X^{} m_p^{}/(m_X^{}+ m_p^{})$ is the reduced mass of the DM and the proton, and $f_p = \sum_{q=u,d,s} f_q^p + (2/9)(1-\sum_{q=u,d,s} f_q^p) \approx 0.468$~\cite{Belanger:2008sj}.

\begin{figure}[t]
\centering
  \includegraphics[width=.49\textwidth]{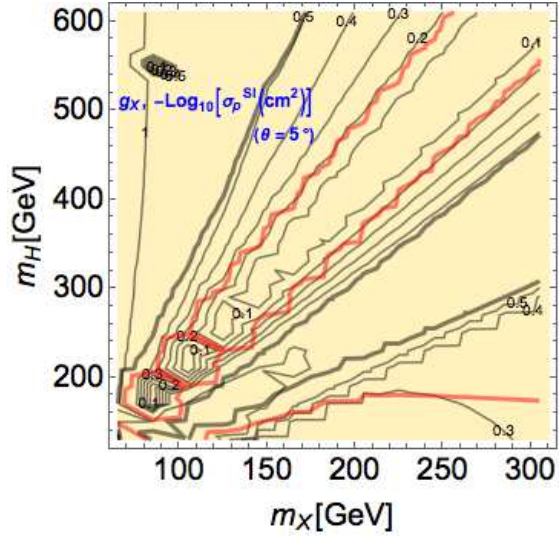}
\caption{
 Contours of $g_X^{}$ (black) and $\sigma_X^{}=10^{-46}~\mathrm{cm}^2$ (red) on the 
 $(m_X^{}, m_H^{})$ for 
 $\Omega_X^{} = \Omega_{\rm obs}$ and $\theta=5^\circ$.
}\label{contour_theta2}
\end{figure}

 The public code {\tt micrOMEGAs 4.3.2}~\cite{Barducci:2016pcb} is used to calculate the thermal VDM relic density $\Omega_X^{} h^2$ and the cross section $\sigma_X^{}$ in this paper.
 Fig.~\ref{contour_theta2} shows the contours of $g_X^{}$ (black) and $\sigma_X^{}=10^{-46}~{\mathrm{cm}}^2$ (red) on the $(m_X^{}, m_H^{})$ plane.
 Here, we assume that the VDM accounts for all the DM energy density {\it i.e.} $\Omega_X^{} = \Omega_{\rm obs}^{}$, and take $\theta=5^\circ$ as a reference point. 
 The observed thermal relic density can be explained at the Higgs poles, $m_X^{}=m_h/2$ and $m_X^{}=m_H^{}/2$~\footnote{
 The Higgs pole of $m_X^{}=m_h/2$ is very narrow.
 We do not perform a fine scanning over this region.
} 
as well as for larger VDM masses $m_X^{}>m_h^{}$ (see Fig.~\ref{fig:200+1_1}).
 In the latter case, the channels $XX \to hh, HH, hH$ contribute to the DM annihilation cross section, which is enhanced as $g_X^{}$ increases (see the figures of $\Omega_X^{}h^2$ plotted as a function of $m_X^{}$ in Refs.~\cite{Baek:2012se,Duch:2015jta}).

\newpage

\end{document}